\documentclass[preprint,aip,pop]{revtex4-1} %
\usepackage{graphicx}
\usepackage{dcolumn}
\usepackage{bm}
\usepackage{pgfplots}
\pgfplotsset{compat=newest}
\usepackage[utf8]{inputenc}
\usepackage[T1]{fontenc}
\usepackage{mathptmx}
\usepackage{subcaption}
\usepackage{etoolbox}
\usepackage{color}
\usepackage{amsmath}

\newcommand{\rf}{r_\mathrm{f}}
\renewcommand{\eqref}[1]{Eq.~(\ref{#1})}

\makeatletter
\def\@email#1#2{%
 \endgroup
 \patchcmd{\titleblock@produce}
  {\frontmatter@RRAPformat}
  {\frontmatter@RRAPformat{\produce@RRAP{*#1\href{mailto:#2}{#2}}}\frontmatter@RRAPformat}
  {}{}
}%
\makeatother
\begin{document}

\preprint{}
\title[Solutions for Rad.~Diffusion in Spherical \& Cylindrical Geometries]{Consistent Solutions of the Radiation Diffusion Equation in Spherical and Cylindrical Geometries}
\author{Ethan Smith}
\affiliation{ 
Los Alamos National Laboratory\\Los Alamos, NM, USA
}%
\author{Evan M. Bursch}%
\affiliation{ 
Department of Applied Physics\\ Columbia University\\ New York City, NY, USA
}%

\author{Ryan G.\ McClarren}
 \email{rmcclarr@nd.edu}
\affiliation{%
Aerospace and Mechanical Engineering\\University of Notre Dame\\Notre Dame, Indiana, USA
}%

\date{\today}

\begin{abstract}
We have extended the radiation diffusion model of Hammer and Rosen to diverging spherical and cylindrical geometries. The effect of curvilinear geometry on the supersonic, expanding wavefront  increases as the internal radius of a spherical or cylindrical shell approaches zero. Small spherical geometries are important for modeling systems at the size scale of ICF capsules, at these scales existing quasi-analytic models for planar geometry significantly disagree with the results of simulation. With this method, the benefits of rapid iteration can be applied to common spherical systems at much smaller length scales. We present comparisons between numerical diffusion solutions and the analytic model to give ranges of applicability for the model.

\end{abstract}

\maketitle

\section{\label{sec:level1}Introduction}

In high energy density conditions, heat transfer via radiative transfer is an important mechanism for understanding the energy balance and evolution of system \cite{2004rahy.book.....C, Drake}. For instance, it is necessary to correctly model radiative transfer in order to get reasonable predictions of the behavior of inertial confinement fusion (ICF) experiments \cite{atzeni2004physics} and other systems of astrophysical and laboratory interest. \cite{Milovich_2021,PhysRevLett.132.065104,MOORE201519}. 

One common phenomenon is the Marshak wave, a heat wave driven by nonlinear radiation diffusion when a radiation source heats an initially cold medium. These systems have been thoroughly studied by several groups, and much work has been done regarding mathematical solutions to the radiation diffusion equation \cite{10.1063/5.0003637,hristov,krief2024unified,cohen2018modeling,lane2013new,Smith:2010ed}. Marshak waves have also been used to understand uncertainty in radiation flow experiments \cite{fryer2016}.   

Hammer and Rosen \cite{10.1063/1.1564599} (hereafter HR) created an approximate model for  radiation propagation in a Marshak wave with an arbitrary drive temperature, such as in the case of indirect laser driven ICF. This model is  for a plane-parallel geometry and takes the form of a single ordinary differential equation (ODE). For systems at the length scale typical of an ICF hohlraum the use of a planar model is satisfactory, with errors estimated at around 10 percent. This error is similar to the errors accrued due to power law assumptions for opacity and internal energy. It has been demonstrated in numerical studies of the fusion capsule after ignition, which is significantly smaller than the hohlraum, that the numerical solution to the radiation diffusion equation in curvilinear geometry differs significantly when compared to the HR method \cite{bellenbaum2023}. Furthermore, it has been noted that the dynamics of a Marshak wave in an ICF capsule determine the burn duration of the fuel \cite{burnRef}. 

In this work we derive an approximate solution to the radiation diffusion equation in spherical and cylindrical coordinates. The major benefit of the HR solution is that it permits an arbitrary boundary temperature instead of requiring either a fixed boundary temperature or a boundary temperature which follows some power law, such as \cite{10.1063/5.0186666,10.1063/5.0088104}. Our model retains this property. 

The model is constructed via an iterative perturbation method which builds up solutions using corrections to the planar model at higher order. This technique is facilitated by the fact that Marshak waves exhibit a very steep gradient near to the heat front. Like HR, we arrive at an expression for the heat front position as an ODE. This quantity is necessary for solving for the temperature profile. With these two quantities, the temperature profile and the effective albedo of the material can be computed semi-analytically. The dynamics of the heat front position are not trivial, especially in curvilinear coordinates; our ODE must be solved numerically. Aside from this one step, the method is analytic, but remains approximate due to some higher order terms being neglected. 

\begin{figure}
    \centering
    \begin{tikzpicture}[scale=1.0]

\def\rO{3.2}      
\def\rOtwo{3.1}   
\def\rft{2.6}     
\def\rzero{1.3}   

\shade[inner color=white, outer color=red!30, draw=black, thick]
  (0,0) circle (\rft);

\shade[inner color=white, outer color=red!50, draw=black, thick]
  (0,0) circle (\rzero);

\draw[thick, gray, dashed] (0,0) circle (\rO);

\draw[->, thick] (0,0) -- ++(0:1.3) node[midway, below] {$r_0$};
\draw[->, thick] (0,0) -- ++(40:\rft) node[midway, right] {$r_\mathrm{f}(t)$};

\end{tikzpicture}
    \caption{Schematic diagram of the considered problem. A temperature boundary is imposed at $r_0$, modeling a hot region radiating out into an initially cold spherical or cylindrical shell. The heat front position $\rf(t)$ moves outward in time.}
    \label{fig:schematic}
\end{figure}
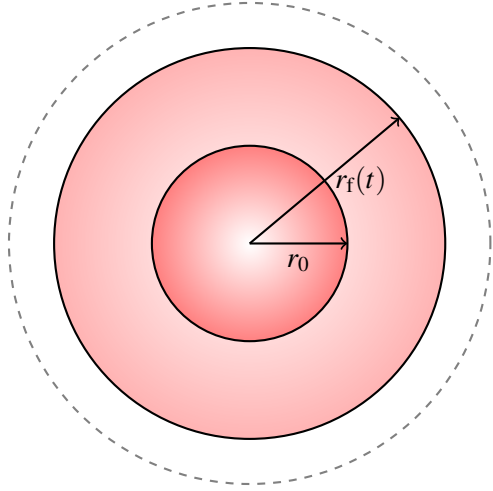

There are several assumptions that have to be made to use this method. Some restrictions on the relative order of terms is challenged by the gradients in curvilinear geometry. This effect is exaggerated as the initial radius shrinks. We also present results only for the case of diverging system, where the internal radius is driven by some radiation drive. A diagram of this system is given in Figure \ref{fig:schematic}. 



\section{Spherical and cylindrical solution to the nonlinear radiation diffusion equation}

We begin with the basic equation for supersonic diffusive radiative transport \cite{10.1063/1.1564599}
\begin{equation}\label{nonlin_diff}
    \rho \frac{\partial e}{\partial t} = \frac{4}{3} \nabla \cdot \left( \frac{1}{K\rho}   \nabla \sigma T^4\right ).
\end{equation}
 Here, $\rho  \left [{\text g}/{\text {cm}^3}\right ]$ is density, $e$ [GJ/g] is specific internal energy, $t$ [\text{ns}] is time, and $T, \left [ \text{HeV} = 100 \,\text{eV} \right] $ is temperature, and $\sigma$ is the Stefan-Boltzmann constant.
We consider curvilinear (either spherical or cylindrical) geometries that vary only in a single radial coordinate, $r$, over a semi-infinite domain given by $r\geq r_0$ where $r_0$ is the minimum radius of the domain. Moreover, we consider the setting where the density, $\rho$, is constant, and that power laws in temperature and density can approximate the internal energy $e = fT^\beta\rho^{-\mu}$ and Rosseland mean opacity $\frac{1}{K} = gT^\alpha\rho^{-\lambda}$. Here $f,$ and $ g,$ are constants that depend on the material. To complete our problem description, our domain is initially at a constant, zero temperature and there is a known, time dependent radiation source at $r_0$ characterized by a temperature, $T(r_0,t) = T_s(t)$.
This relation serves as a boundary condition at $r_0$. 

Additionally, we impose that beyond the wavefront at a position $\rf(t)$, the medium is cold and the temperature is zero, i.e., $T(r > \rf,t) = 0$. Due to the nonlinear nature of \eqref{nonlin_diff}, at the wavefront, the flux of radiation is also zero implying that the derivative of $T$ with respect to $r$ is also zero, i.e., 
\[ \left.\frac{\partial T}{\partial r}\right|_{\rf} = 0.\]
For a further discussion of these boundary conditions see \cite{2004rahy.book.....C}.

With these power law assumptions and some simplification, we arrive at a form parameterized by geometry where $d = 0,1,2$ corresponds to planar, cylindrical, and spherical geometries, respectively:
\begin{equation}\label{eq2}
    \frac{\partial T^\beta}{\partial t} = \frac{C}{r^d}\frac{\partial}{\partial r} \left( r^d \frac{\partial }{\partial r}\left(T^{4+\alpha} \right) \right).
\end{equation}
Here our constants have been gathered as
\begin{equation}
C = \frac{4}{4+\alpha}\frac{4}{3}\frac{1}{f}g\rho^{\mu-2-\lambda}
.
\end{equation}


Now we introduce terms that will allow us to nondimensionalize,  which will simplify our solution. The first is the surface drive, $H(t)$, given by
\begin{equation}
    H(t) = T_s^{4+\alpha}(t).
\end{equation} 
We also define for convenience
\begin{equation}
    \zeta(r,t) = \frac{T^{4+\alpha}}{T_s^{4+\alpha}}
\end{equation}
as a dimensionless temperature-like quantity. With these definitions \eqref{eq2} becomes
\begin{equation}\label{eq:diff_nondim}
    \frac{\partial}{\partial t} \left( H^\epsilon \zeta ^\epsilon \right) = CH\frac{1}{r^d} \frac{\partial}{\partial r}\left( r^d \frac{\partial \zeta}{\partial r}\right) ,
\end{equation}
where  $\epsilon = \frac{\beta}{4+\alpha}$. In many materials in the HED regime, $\epsilon \ll 1$, and we use this fact to derive approximate solutions to \eqref{eq:diff_nondim}.

We also define a dimensionless spatial variable, 
\begin{equation}\label{y}
    y = \frac{r-r_0}{\rf-r_0},
\end{equation}
where $\rf$ is the time-dependent distance the heat front has traveled from the surface at $r_0$; with this definition, the heat wave extends from $y=0$ at $r_0$ to $y=1$ at $\rf$. 
We also adopt the non-dimensional boundary conditions, which are the original boundary conditions rewritten using the nondimensional temperature, 
\begin{equation}\label{BC}
    \zeta(0,t) = 1,\textrm{ } \zeta(1,t) = 0 \textrm{, and } \left.\frac{\partial \zeta}{\partial y}\right|_{y=1} = 0.
\end{equation}
The time derivative transforms as
\begin{equation}\label{eq:r_deriv}
    \left.\frac{\partial}{\partial t}\right|_r = \left.\frac{\partial}{\partial t}\right|_y + 
    \left.\frac{\partial y}{\partial t}\right|_r\frac{\partial}{\partial y},
\end{equation}
leading to
\begin{equation} \label{eq:y_deriv}
    \left.\frac{\partial y}{\partial t}\right|_r = \frac{1}{\rf-r_0}\frac{\partial}{\partial t}(r-r_0) + (r-r_0)\frac{\partial}{\partial t}\left(\frac{1}{\rf-r_0}\right).
\end{equation}
The first term in \eqref{eq:y_deriv} evaluates to zero and the second term leads to the following expression for the derivative
\begin{equation}
    \left.\frac{\partial}{\partial t}\right|_r = \left.\frac{\partial}{\partial t}\right|_y - \frac{r-r_0}{(\rf-r_0)^2}\frac{\partial \rf}{\partial t} \frac{\partial }{\partial y}.
\end{equation}
We substitute in our evaluated time derivative into \eqref{eq:diff_nondim} 
\begin{equation}
    \left( \left.\frac{\partial}{\partial t}\right|_y - \frac{y}{\rf-r_0}\frac{\partial \rf}{\partial t} \frac{\partial }{\partial y}\right)H^\epsilon\zeta^\epsilon =CH\frac{1}{r^d} \frac{\partial}{\partial r}\left( r^d \frac{\partial \zeta}{\partial r}\right).
\end{equation}
Expanding this operator applied to $H^\epsilon\zeta^\epsilon$ gives
\begin{equation}\label{eq:diffusion_t}
    \left(\frac{\partial H^\epsilon}{\partial t} \right)\zeta^\epsilon + H^\epsilon\frac{\partial\zeta^\epsilon}{\partial t} - H^\epsilon\frac{y}{\rf-r_0}\frac{\partial \rf}{\partial t}\frac{\partial \zeta^\epsilon}{\partial y} = CH\frac{1}{r^d} \frac{\partial}{\partial r}\left( r^d \frac{\partial \zeta}{\partial r}\right).
\end{equation}
This is a general geometry version of Eq.~(11) from \cite{10.1063/1.1564599}; these can be seen by setting $d=r_0=0$.

Now we change to  time variable, $s(t)$, given by
\begin{equation}\label{eq:s}
    s(t) = \int_0^t \frac{CH^{1-\epsilon}(\tau)}{(\rf(\tau)-r_0)^2} d\tau.
    \end{equation}  
    With this definition, time derivatives transform via
    \begin{equation}
\frac{\partial s}{\partial t} = \frac{CH^{1-\epsilon}(t)}{(\rf(t)-r_0)^2},
    \text{\quad and \quad } \frac{\partial }{\partial t} = \frac{\partial s}{\partial t}\frac{\partial }{\partial s}.
\end{equation}
Applying this transformation to \eqref{eq:diffusion_t} leads to
\begin{equation}
    \left(\frac{CH^{1-\epsilon}}{(\rf-r_0)^2} \frac{\partial H^\epsilon}{\partial s}\right)\zeta^\epsilon + \frac{CH}{(\rf-r_0)^2}\frac{\partial \zeta^\epsilon}{\partial s} - \frac{CH}{(\rf-r_0)^2}\frac{y}{\rf-r_0}\frac{\partial \rf}{\partial s}\frac{\partial \zeta^\epsilon}{\partial y}=CH\frac{1}{r^d} \frac{\partial}{\partial r}\left( r^d \frac{\partial \zeta}{\partial r}\right).
\end{equation}
Dividing through by $CH$ and multiplying through by $(\rf-r_0)^2$ gives 
\begin{equation}
    \left(H^{-\epsilon} \frac{\partial H^\epsilon}{\partial s} \right)\zeta^\epsilon + \frac{\partial \zeta^\epsilon}{\partial s} - \frac{y}{\rf-r_0}\frac{\partial \rf}{\partial s}\frac{\partial \zeta^\epsilon}{\partial y} = \frac{(\rf-r_0)^2}{r^d}\frac{\partial}{\partial r}\left( r^d \frac{\partial \zeta}{\partial r}\right).
\end{equation}
Finally, we transform the spatial derivatives in $r$ according to  \eqref{y}
\begin{equation}
    \frac{\partial}{\partial r} = \frac{\partial y}{\partial r}\frac{\partial }{\partial y} = \frac{1}{(\rf-r_0)}\frac{\partial}{\partial y}.
\end{equation}
After simplification, we arrive at
\begin{equation}
    \frac{\epsilon}{H} \frac{\partial H}{\partial s}\zeta^\epsilon + \frac{\partial \zeta^\epsilon}{\partial s} - \frac{y}{\rf-r_0}\frac{\partial \rf}{\partial s}\frac{\partial \zeta^\epsilon}{\partial y} = \frac{\partial^2 \zeta}{\partial y^2} + \frac{d\cdot(\rf-r_0)}{r(y)}\frac{\partial \zeta}{\partial y}.
\end{equation}

Following \cite{10.1063/1.1564599} we add a term to both sides to alleviate issues whenever $\rf = r_0$ and substitute expressions for $r(y)$:

\begin{equation}
    \frac{\epsilon}{H} \frac{\partial H}{\partial s}\zeta^\epsilon + \frac{\partial \zeta^\epsilon}{\partial s} + \frac{1-y}{\rf-r_0}\frac{\partial \rf}{\partial s}\frac{\partial \zeta^\epsilon}{\partial y} = \frac{\partial^2 \zeta}{\partial y^2} + \frac{1}{(\rf-r_0)}\frac{\partial \rf}{\partial s}\frac{\partial\zeta^\epsilon}{\partial y}+ \frac{d(\rf-r_0)}{y(\rf - r_0) + r_0}\frac{\partial \zeta}{\partial y}.
\end{equation}
For brevity we introduce $\gamma = (\rf-r_0)$ and move the last term on the right hand side to the left hand side,
\begin{equation}\label{ours}
\frac{\epsilon}{H} \frac{\partial H}{\partial s}\zeta^\epsilon + \frac{\partial \zeta^\epsilon}{\partial s} + \frac{1-y}{\gamma}\frac{\partial \rf}{\partial s}\frac{\partial \zeta^\epsilon}{\partial y} - \frac{d\cdot\gamma}{y\gamma + r_0}\frac{\partial \zeta}{\partial y} = \frac{\partial}{\partial y}\left(\frac{\partial \zeta}{\partial y} + \frac{1}{\gamma}\frac{\partial \rf}{\partial s}\zeta^\epsilon\right). 
\end{equation}

\subsection{Solution procedure for transformed radiation diffusion equation}
Equation \eqref{ours}  is similar to \cite[Eq.~(13)]{10.1063/1.1564599}, with the inclusion of a correction term which depends on geometry. As argued in \cite{10.1063/1.1564599}, the left hand side of \eqref{ours} will be small away from the wavefront. The same arguments will hold in the curvilinear case provided the initial radius, $r_0$, is not too small (a consideration that we will return to later). Therefore, we follow the procedure of setting the left hand side equal to zero and solving the right hand side to get a lowest order approximation for the pseudo-temperature profile $\zeta$ and the heat front position $\rf$. The resulting equation that we must solve to get this lowest order approximation is
\begin{equation}\label{RHS0}
   \frac{\partial}{\partial y}\left(\frac{\partial \zeta}{\partial y} + \frac{1}{\gamma}\frac{\partial \rf}{\partial s}\zeta^\epsilon\right) 
   =0.
\end{equation}
Next we integrate \eqref{RHS0} over $y$ from $y$ to 1 and apply the boundary conditions at $y=1$ to get a differential equation that has the solution
\begin{equation}\label{theirs}
    \zeta = \left(1 - (\epsilon-1)\frac{1}{\gamma}\frac{\partial \rf}{\partial s}y \right)^{\frac{1}{1-\epsilon}}.
\end{equation}
Imposing the boundary condition $\zeta(1,s)=0$, we arrive at 
\begin{equation}\label{eq:bc_drf}
    1 = (\epsilon-1)\frac{1}{\gamma}\frac{\partial \rf}{\partial s}.
\end{equation}
From this equation we can conclude that the leading order approximate solution to \eqref{ours} is
\begin{equation}
  \zeta = (1 - y)^{\frac{1}{1-\epsilon}} = (1-y) + O(\epsilon).
\end{equation}
We call this approximate solution $\zeta_0 \equiv 1-y$.

We use Picard iteration to improve on the approximation $\zeta_0$. To accomplish this we substitute $\zeta_0$ into the lefthand side of \eqref{ours} to get an updated version of \eqref{RHS0} with a non-zero righthand side. 
This technique quickly yields results; our final approximate solution is accurate through first order in $\epsilon$. The process begins by selectively substituting the solution, $\zeta_0$ into the left hand side of \eqref{ours} to yield
\begin{equation}
    \epsilon\frac{1}{H}\frac{\partial H}{\partial s}\zeta_0^\epsilon - \frac{\epsilon}{\gamma}\frac{\partial \rf}{\partial s}\zeta_0^\epsilon -\frac{d\cdot\gamma}{y\gamma + r_0}\frac{\partial \zeta_0}{\partial y}= \frac{\partial}{\partial y}\left(\frac{\partial \zeta}{\partial y} + \frac{1}{\gamma}\frac{\partial \rf}{\partial s}\zeta^\epsilon\right).
\end{equation}
Here we have used the fact that $\zeta_0$ does not depend on $s$ and that 
\[
(1-y) \frac{\partial}{\partial y}(1-y)^\epsilon = - \epsilon (1-y)^\epsilon = -\epsilon \zeta_0.
\]
We integrate both sides over $y$ from $y$ to $1$. In the process we use the fact that  
\begin{equation}
    \int_y^1\zeta_0^\epsilon(\hat{y}) \,d\hat{y} = \frac{(1-y)^{\epsilon +1}}{\epsilon +1} = 1 - y + O(\epsilon),
\end{equation} and ${\partial \zeta_0}/{\partial y} = -1$. This results in 
\begin{equation}
    \epsilon (y-1)\left(\frac{1}{H}\frac{\partial H}{\partial s} - \frac{1}{\gamma}\frac{\partial \rf}{\partial s}\right)-d\cdot\gamma\int_y^1{\frac{1}{y\gamma + r_0}d\hat{y}}= \frac{\partial \zeta}{\partial y} + \frac{1}{\gamma}\frac{\partial \rf}{\partial s}\zeta^\epsilon.    
\end{equation}
Evaluating the integral yields a logarithm:
\begin{equation}
    \epsilon (y-1)\left(\frac{1}{H}\frac{\partial H}{\partial s} - \frac{1}{\gamma}\frac{\partial \rf}{\partial s}\right)+d\,\log\left(\frac{\gamma y + r_0}{\gamma + r_0}\right) = \frac{\partial \zeta}{\partial y} + \frac{1}{\gamma}\frac{\partial \rf}{\partial s}\zeta^\epsilon.    
\end{equation}
This equation has an analogue in previous work \cite[Eq.~(19)]{10.1063/1.1564599} with an added logarithmic term for non-planar geometry. We also note that as $r_0 \rightarrow \infty$ we get the planar result because the logarithm and $\gamma^{-1}$ terms limit to zero.

We divide through by $\zeta^\epsilon$ since the chain rule implies
\begin{equation}
    \zeta^{-\epsilon}\frac{\partial \zeta}{\partial y} = \frac{1}{1-\epsilon}\frac{\partial \zeta^{1-\epsilon}}{\partial y}.
\end{equation}
Also, we set  $\zeta^\epsilon \approx 1$ in all denominators on the left hand side and selectively apply $\frac{1}{\gamma}\frac{\partial \rf}{\partial s}=1 + O(\epsilon)$ from \eqref{eq:bc_drf} to just the left hand side to get the simplified form:
\begin{equation}
        \epsilon (y-1)\left(\frac{1}{H}\frac{\partial H}{\partial s} - 1\right)+d\,\log\left(\frac{\gamma y + r_0}{\gamma + r_0}\right)= \frac{1}{1-\epsilon}\frac{\partial \zeta^{1-\epsilon}}{\partial y} + \frac{1}{\gamma}\frac{\partial \rf}{\partial s}.  
\end{equation}

Now we integrate over $y$ from $y=0$ to $y$ to yield a solution for $\zeta$:
\begin{equation}
  \zeta =  \left((1-\epsilon)\left[ \epsilon\left(\frac{y^2}{2}-y\right)\left(\frac{1}{H}\frac{\partial H}{\partial s} - 1\right) + d\,\left( -y +  \frac{r_0}{\gamma}\log\frac{\gamma y}{r_0} + 1  \right) - y\frac{1}{\gamma}\frac{\partial \rf}{\partial s}\right] + 1 \right)^{\frac{1}{1-\epsilon}}
\end{equation}
This equation has a term proportional to $d$ that is a geometric correction for non-planar geometry.
We call this correction term
\begin{equation}
    \psi = -y d + \frac{r_0 d}{\gamma}\log\left(\frac{\gamma y}{r_0} + 1\right),
\end{equation}
and try to rearrange the previous equation to something more like \cite[Eq.~(20)]{10.1063/1.1564599}. 
 We also note that $\psi$ goes to zero as $r_0, \rf \rightarrow \infty$; this is expected because the effects of curvilinear geometry should also vanish as the radius goes to infinity. Substituting for $\psi$ and omitting terms of $O(\epsilon^2)$, we get
 \begin{equation}\label{eq:zeta_rfds}
   \left((\epsilon-\epsilon^2) \left(y-\frac{y^2}{2}\right)\left(1 - \frac{1}{H}\frac{\partial H}{\partial s}\right) + (1-\epsilon)\psi - (1-\epsilon)y\frac{1}{\gamma}\frac{\partial \rf}{\partial s} + 1 \right)^{\frac{1}{1-\epsilon}}= \zeta
\end{equation}

We continue by applying the boundary condition $\zeta(1,s) = 0$ and solving for $(1-\epsilon)\frac{1}{\gamma}\frac{\partial \rf}{\partial s}$. 
We arrive at an ODE for $\rf$:
\begin{equation}\label{rfODE}
  \frac{\partial \rf}{\partial s} = \frac{\gamma}{1-\epsilon}\left(\epsilon \left(\frac{1}{2}\right)\left(1 - \frac{1}{H}\frac{\partial H}{\partial s}\right) + (1-\epsilon)\left( -d+ \frac{d \cdot r_0}{\gamma}\textrm{ln}\left(\frac{\gamma}{r_0} + 1\right)\right) + 1 \right).
\end{equation}

We substitute the value of $\frac{\partial \rf}{\partial s}$ in  \eqref{eq:zeta_rfds}, leading to the solution for $\zeta$ to $O(\epsilon)$ given by 
\begin{multline}
    \zeta=\left(\epsilon \left(y-\frac{y^2}{2}\right)\left(1 - \frac{1}{H}\frac{\partial H}{\partial s}\right) + (1-\epsilon)\psi - \right. \\ \left. y\left[ \epsilon \left(\frac{1}{2}\right)\left(1 - \frac{1}{H}\frac{\partial H}{\partial s}\right) + (1-\epsilon)\left( -d+ \frac{d\cdot r_0}{\gamma}\textrm{ln}\left(\frac{\gamma}{r_0} + 1\right)\right) + 1 \right] + 1 \right)^{\frac{1}{1-\epsilon}}.
\end{multline}
 We can show that the boundary condition $\zeta(1,s) = 0$ is still satisfied because $(y-y^2/2)$ evaluates to one-half at $y=1$ and the term involving $\psi(1,s)$ is exactly canceled as well. Though we have a solution for $\zeta$, to evaluate this for a given driving temperature profile, we need to compute $s$, but $s$ is defined via an integral of $\rf$. Therefore, we need an expression for the wavefront position as a function of time. In planar geometry this can be determined via a closed form solution. In the general case we will have an ODE to solve. Recall that $\frac{\partial s}{\partial t} = \frac{CH^{1-\epsilon}}{\gamma^2}$ and $\frac{\partial }{\partial t} = \frac{\partial s}{\partial t}\frac{\partial }{\partial s}$. Using these facts we arrive at a form of \eqref{rfODE} in the typical time coordinate $t$
\begin{equation}\label{ode_for_rf}
  \frac{1}{1-\epsilon}\left( \frac{\epsilon}{2}\left(1 - \frac{\gamma^2}{CH^{2-\epsilon}}\frac{\partial H}{\partial t}\right) + (1-\epsilon)\left( -d+ \frac{d \cdot r_0}{\gamma}\textrm{ln}\left(\frac{\gamma}{r_0} + 1\right)\right) + 1 \right)= \frac{\gamma}{CH^{1-\epsilon}}\frac{\partial \rf}{\partial t}.
\end{equation}
We will numerically integrate \eqref{ode_for_rf} using an ODE solver starting at $t=0$ where $\rf(0) = r_0$ and $H(0)$ and its derivative are known. 
After the numerical solution of $\rf(t)$, we then can compute the non-dimensional time $s(t)$ using \eqref{eq:s} to then evaluate $\partial \rf / \partial s$. This derivative is then used in \eqref{eq:zeta_rfds} to evaluate the $\zeta$ profile.

In constructing the solution with a Picard iterative procedure, we have made the assumption that the terms on the left side of \eqref{ours} are small relative to the right hand side. This essentially places a minimum value of the inner radius of the domain, $r_0$, since the geometric correction grows as $r_0$ decreases. At a glance, it would seem appropriate to place this term on the other side of the equation before performing Picard iteration. Alas, our forays down this path have been fruitless to date.

We also note that the parameter $\epsilon$, the smallness of which is also used in our analysis, is itself a material property. As a result we expect that the geometric effects can differ between materials. Furthermore, the relative magnitude of the geometrical correction terms change over time and space as the wavefront propagates to larger radii.

\section{Results of this method with various materials and geometries.}

We compare the results of this semi-analytic method to a numerical solution of the PDE in \eqref{nonlin_diff} in curvilinear geometry using the finite volume method in space \cite[Chap.~18]{PythonBook} and  backward differentiation formula (BDF) time integration as implemented in \texttt{SciPy} \cite{2020SciPy-NMeth}. We use automatic time step control and 300 spatial cells over the domain. The domain size varies by material. In general we expect the numerical diffusion model to be an accurate solution. This improved accuracy comes at a significantly increased computational cost.

There are two material properties which uniquely define a material in the context of this semi-analytic method, namely $C$ and $\epsilon$. The selection of materials in this study was selected from Cohen \cite{PhysRevResearch.2.023007} to well span the space of these two parameters using  realistic materials. 

First, we present the various materials used in the study by tabulating their values of $C$ and $\epsilon$ in Table \ref{tableCeps}.
\begin{table}
\caption{Materials Considered \cite{10.1063/1.1564599,PhysRevResearch.2.023007}}
\centering
\begin{tabular}{l|cccc} 
Material & C & $\epsilon$ & $\alpha$ & $\rho$ [g/cm$^3$]\\ \hline
$\text{C}_{15}\text{H}_{20}\text{O}_{6}$ & $7.896\times 10^{-5}$ & $0.1011$ & 5.29 & 0.0575 \\
$\text{C}_6\text{H}_{12}$       & $9.998 \times 10^{-4}$ & $0.1433$ & 2.98 & 0.05 \\
$\text{SiO}_2$        & $7.503 \times 10^{-4}$ & $0.2050$ & 2.0 & 0.03\\
Au        & $1.123 \times 10^{-5}$ & $0.2909$ & 1.5 & 0.2\\ 
\end{tabular}
\label{tableCeps}
\end{table}

The same boundary temperature profile, plotted here, was used for all simulations.
\begin{figure}
    \centering
    \begin{tikzpicture}
  \begin{axis}[
    width=6in,
    height=3.71in,
    xlabel={Time (ns)},
    ylabel={$T_s$ (HeV)},
    xmin=0, xmax=3.0,
    ymin=1.0, ymax=2.1,
    xtick={0,1,2,3},
    ytick={0,0.5,1,1.2,2,2.5},
    grid=major,
    axis lines=left,
    x axis line style={very thick},
    y axis line style={very thick},
  ]
  \addplot [blue, very thick] coordinates {
    (0, 1.2)
    (1, 1.2)
    (2, 2.0)
    (3, 2.0)
  };
  \end{axis}
\end{tikzpicture}
    \caption{Temperature boundary condition for the test problem.}
    \label{fig:wallTemp}
\end{figure}
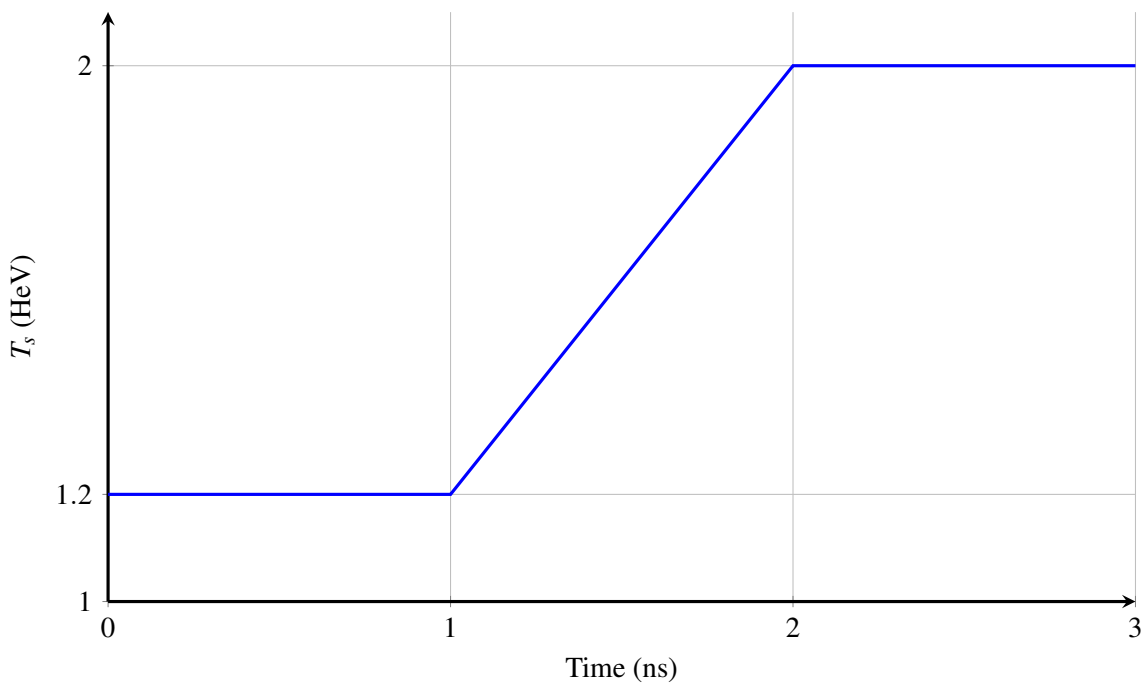
To form this comparison, a system where the left boundary is driven by the same ramp profile drive as in the original presentation of the planar model \cite{10.1063/1.1564599}, this is shown in Figure \ref{fig:wallTemp}. For each material, the comparison was made at several starting radii. Smaller initial radii increased the effect of the curvilinear geometry, and in general the approximate solution performs worse as this parameter is decreased. 
\subsection{Comparison of wavefront position}
\begin{figure}
    \centering
    \begin{subfigure}{0.45\textwidth}
    \includegraphics[width=\linewidth]{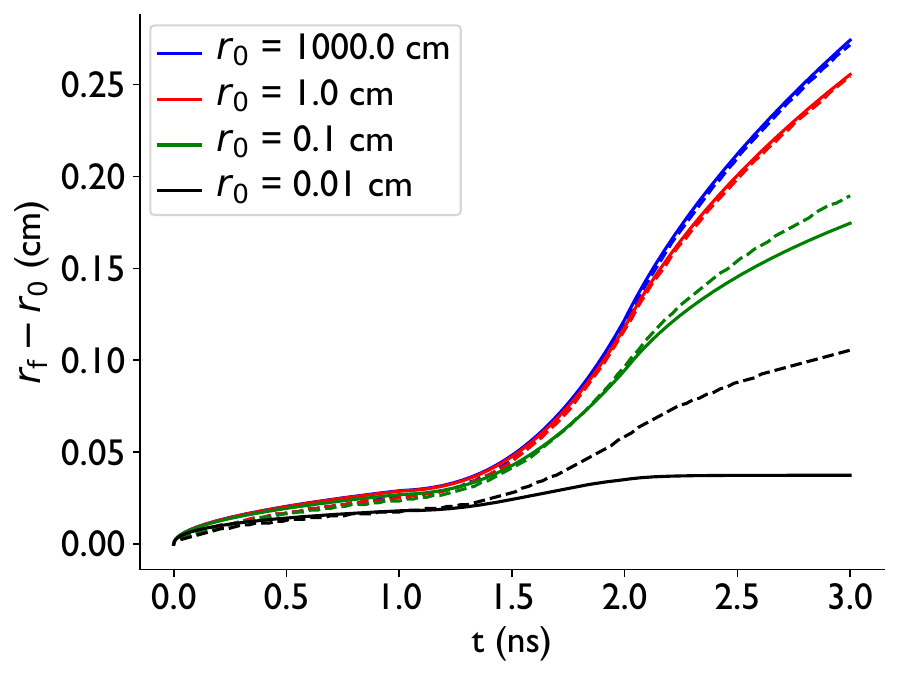}
        \caption{C$_{15}$H$_{20}$O$_6$, $\epsilon = 0.10111$}
       
    \end{subfigure}
    \hfill
    \begin{subfigure}{0.45\textwidth}
     \includegraphics[width=\linewidth]{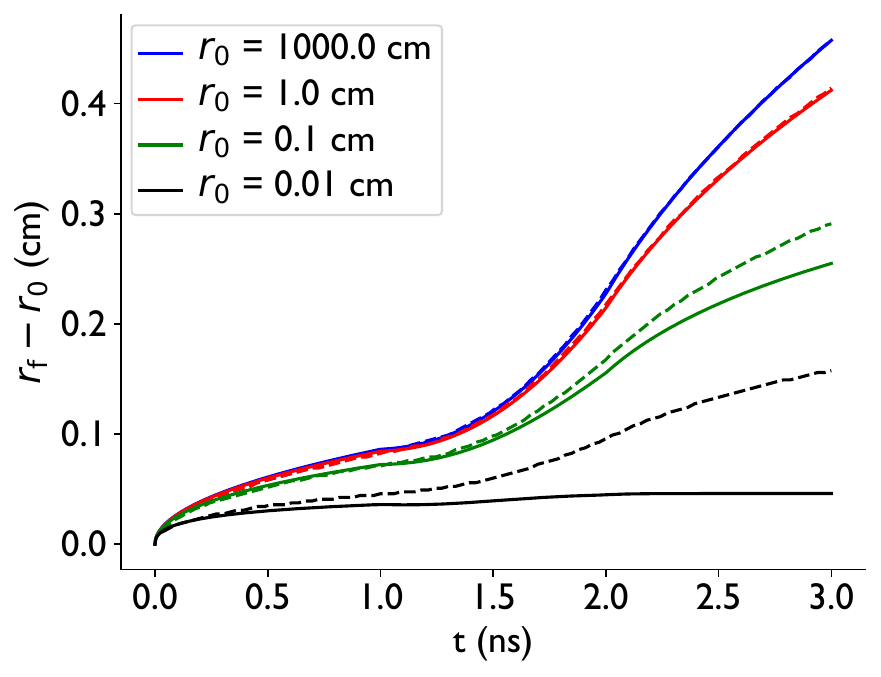}
        \caption{C$_{6}$H$_{12}$, $\epsilon = 0.1433$}
        
    \end{subfigure}

    \vspace{5mm} 
    \begin{subfigure}{0.45\textwidth}
        \includegraphics[width=\linewidth]{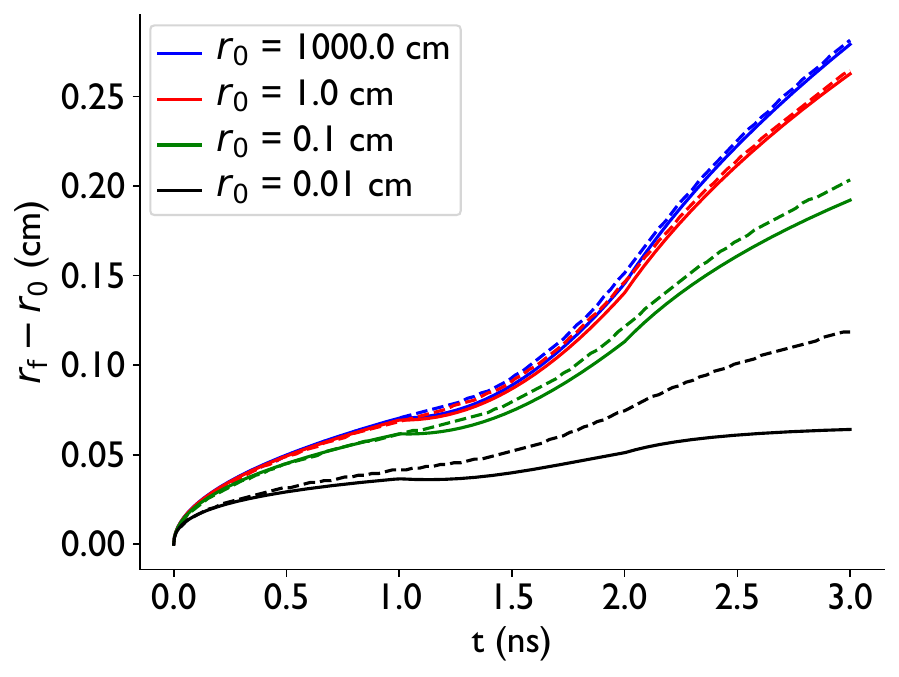}
        \caption{SiO$_2$, $\epsilon = 0.2050$}
    \end{subfigure}
    \hfill
    \begin{subfigure}{0.45\textwidth}
        \includegraphics[width=\linewidth]{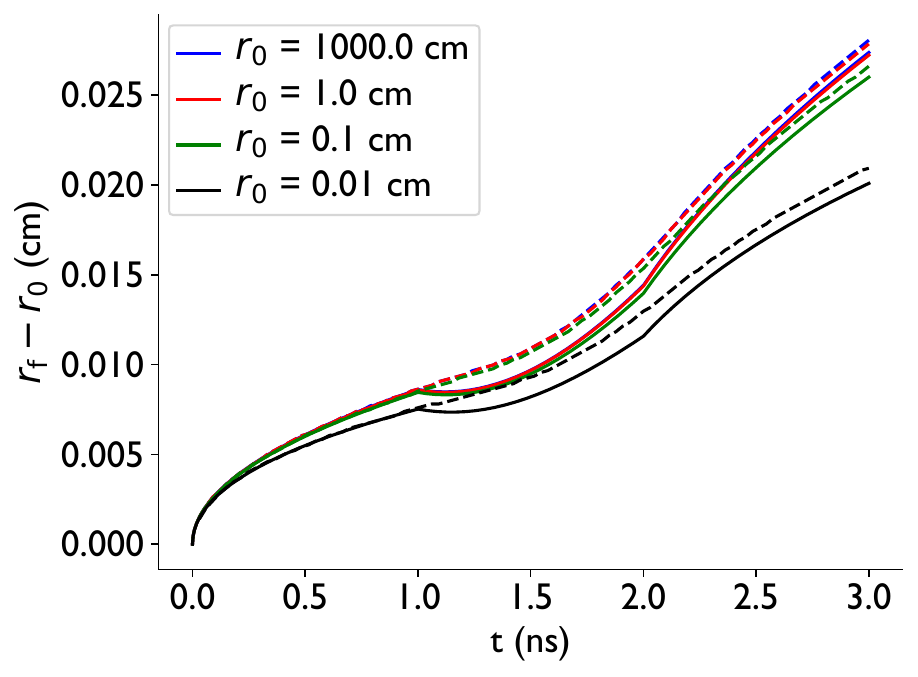}
        \caption{Au, $\epsilon = 0.2909$}
    \end{subfigure}
    
    \caption{Our approximate model, solid line, compared to a numerical diffusion solution, dashed line, for several materials in spherical geometry for a variety of initial radii.}
    \label{fig:SpherePosition}
\end{figure}

We plot the heat front position from our model and from the numerical diffusion solution for four different materials in Figure \ref{fig:SpherePosition}. In the figure, the inner radius of the sphere, $r_0$, takes values of 0.01, 0.1, 1, and 1000 cm. The 1000 cm case is comparable to the planar solution, so the difference between this solution and the other inner radii is a measure of the effect of curvilinear geometry. From the figure we see that neglecting geometric effects would lead to large errors in the heat front position. For instance, in SiO$_2$ the $r_0=0.01$ cm wavefront is about 150\% slower than the $r_0 = 1000$ cm case. This supports the findings of \cite{bellenbaum2023} that a planar model is not sufficient to model spherical geometry at these scales. 

From the figure we see that the smaller the value of $\epsilon$ the larger the magnitude of the geometric effects. For instance, C$_{15}$H$_{20}$O$_6$ with the smallest value of $\epsilon$ has the largest effect and gold, with the largest value of $\epsilon$ has the smallest effect.  Furthermore, for larger values of $\epsilon$ our model is accurate over the range of inner radii. For small values of $\epsilon$ the model slows down the wave too much at $r_0=0.01$ cm. In all of the materials, the error in the final wavefront position at 3 ns is about 10\% or less for $r_0 \geq 0.1$ cm.


\begin{figure}
    \centering
    \begin{subfigure}{0.49\textwidth}
        \includegraphics[width=\linewidth]
        {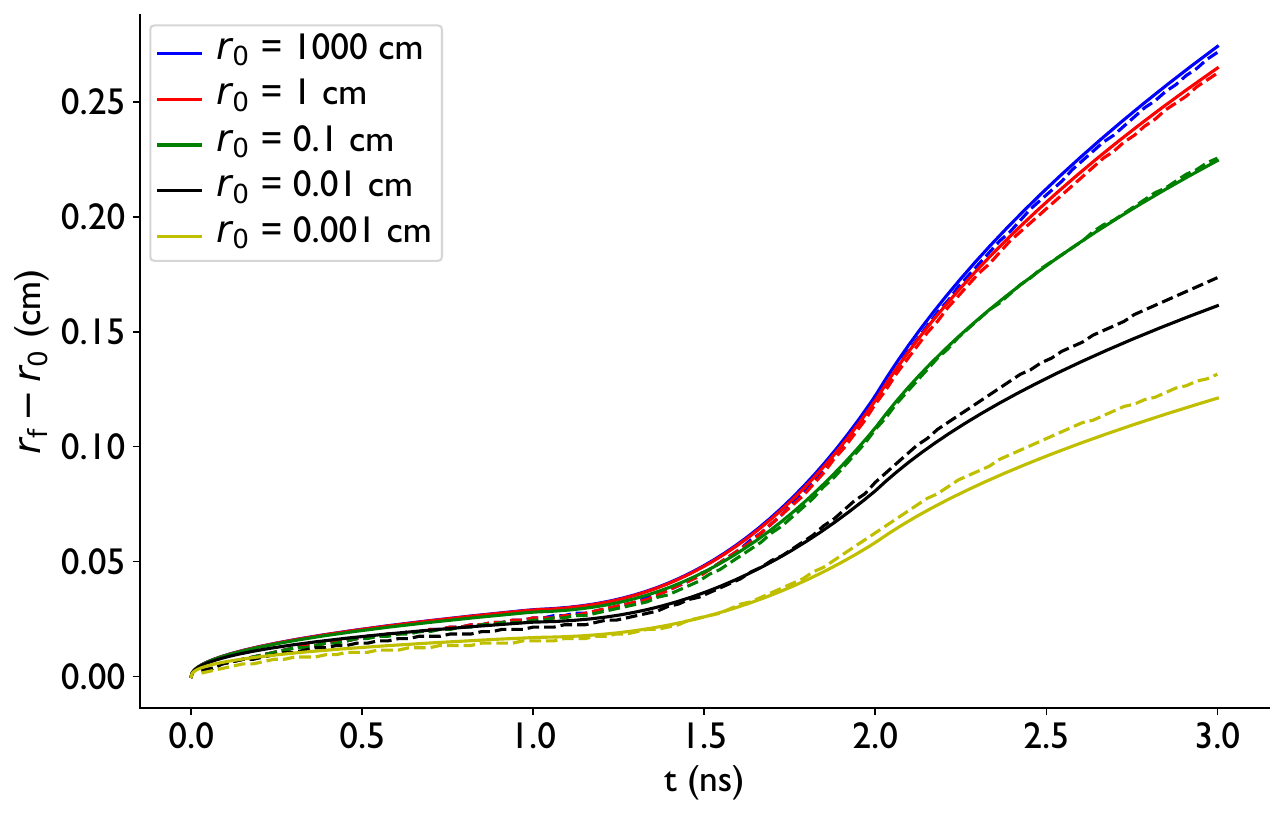}
        \caption{C$_{15}$H$_{20}$O$_6$, $\epsilon = 0.10111$}
    \end{subfigure}
    \hfill
    \begin{subfigure}{0.49\textwidth}
    \includegraphics[width=\linewidth]{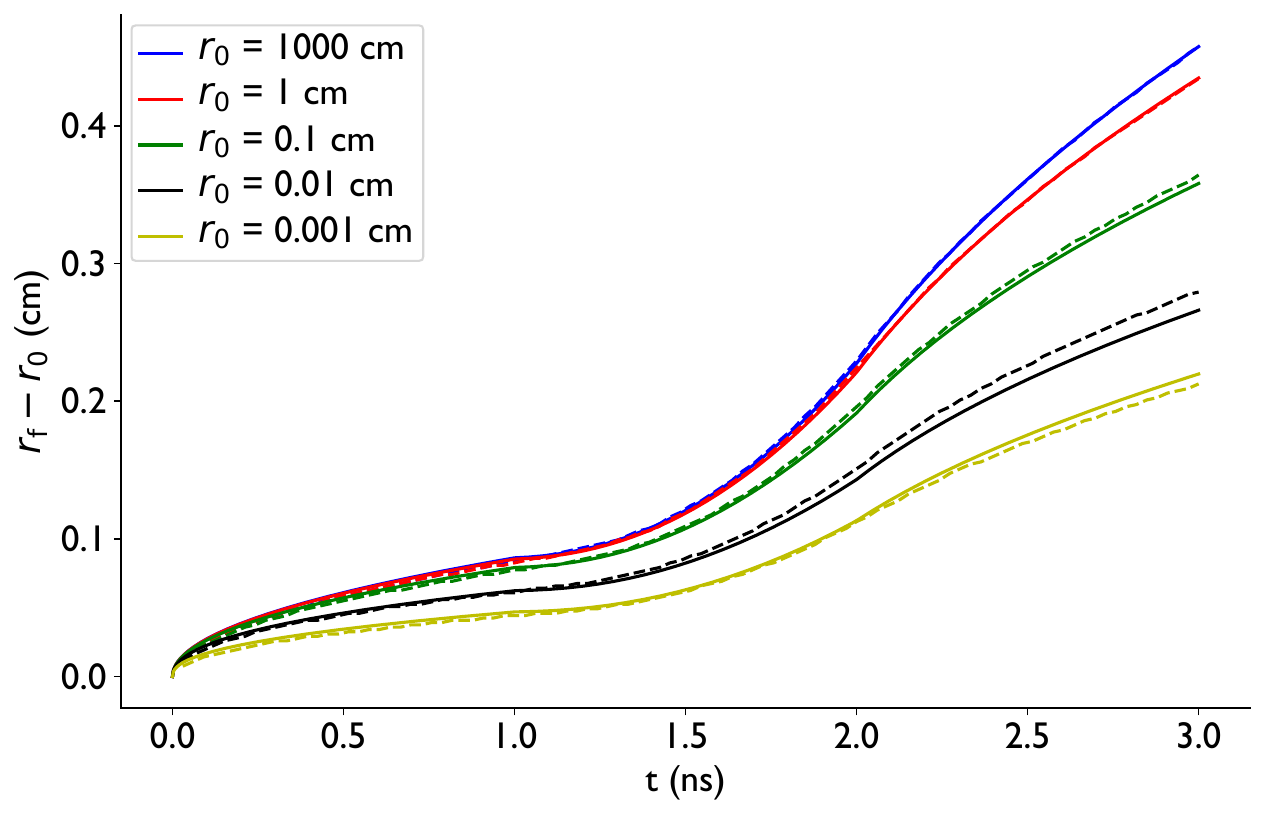}
        \caption{C$_{6}$H$_{12}$, $\epsilon = 0.1433$}
        
    \end{subfigure}

    \vspace{5mm} 
    \begin{subfigure}{0.49\textwidth}
        \includegraphics[width=\linewidth]{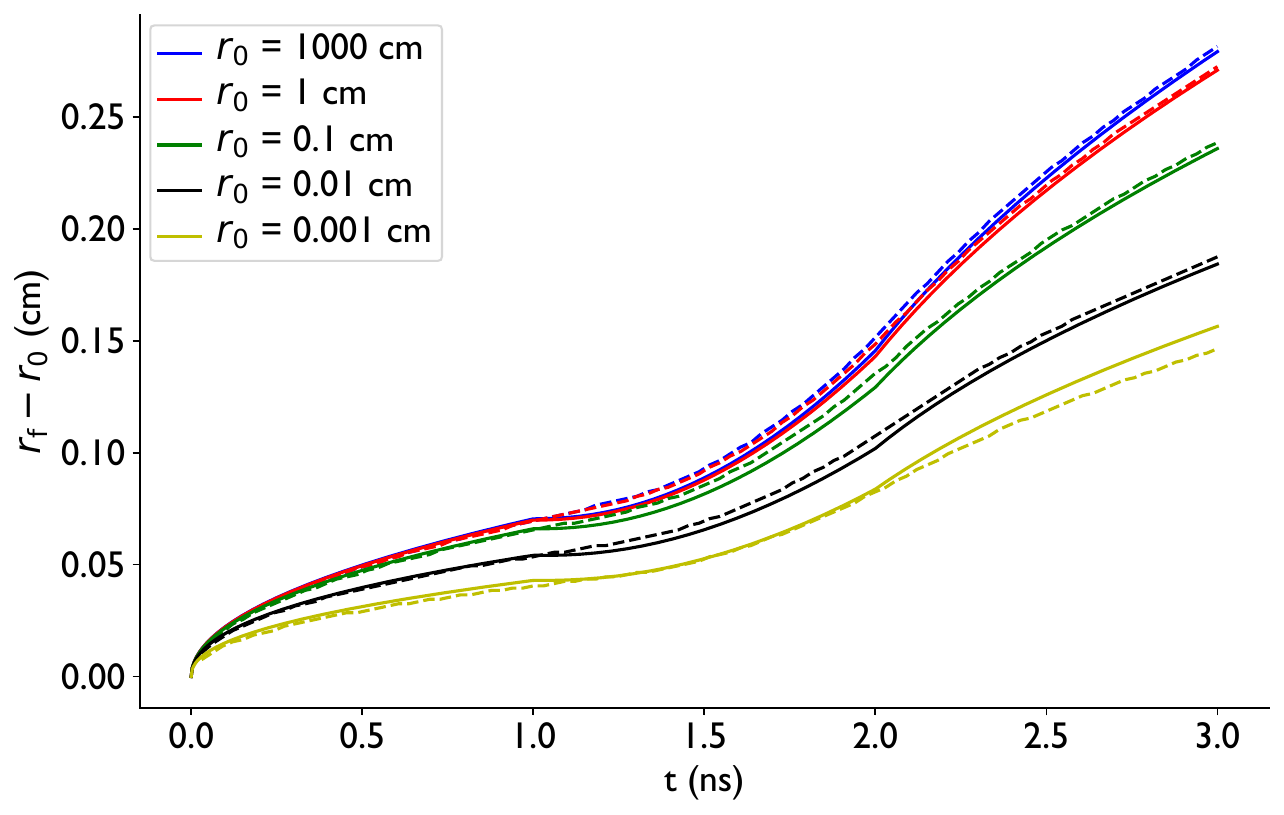}
        \caption{SiO$_2$, $\epsilon = 0.2050$}
        
    \end{subfigure}
    \hfill
    \begin{subfigure}{0.49\textwidth}
        \includegraphics[width=\linewidth]{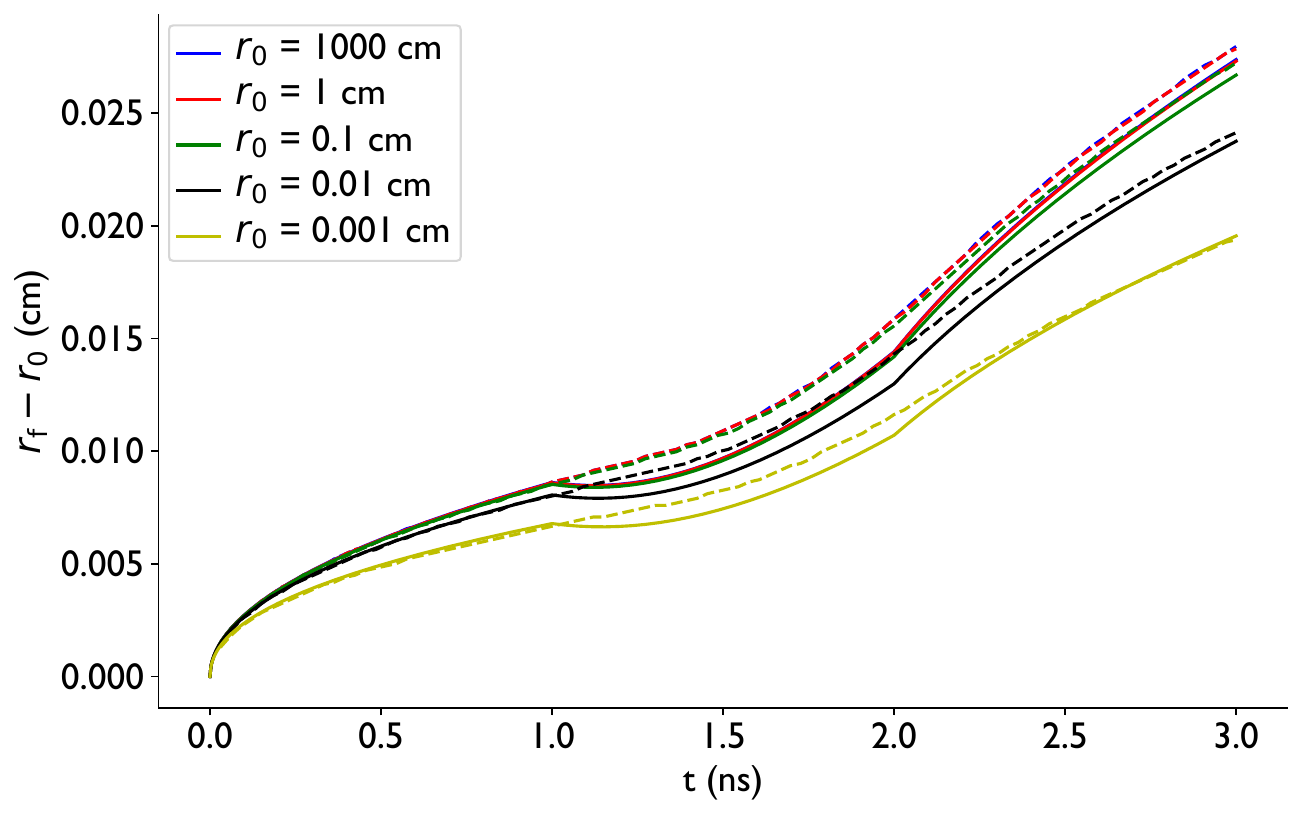}
        \caption{Gold, $\epsilon = 0.2909$}
    \end{subfigure}
    
    \caption{Our approximate model (solid line) compared to a numerical diffusion solution, (dashed line) for several materials in cylindrical geometry for a variety of initial radii. The model shows better agreement in the cylindrical case than the spherical case.}
    \label{cylinPosition}
\end{figure}

Figure \ref{cylinPosition} presents the same calculations but in cylindrical geometry. From this figure we see that geometric effects are reduced in the cylindrical case. This is likely due to the lower magnitude of the correction term $\psi$, which is assumed to be small in both models and is in fact a function of initial radius. In spherical geometry this $\psi$ is twice as large as in cylindrical geometry for a given value of $r_0$.

We note all models have a non-smooth change in the solution whenever the boundary condition changes from being held at a constant value to beginning the ramp at 1 ns. The effect is largest in spherical geometry and when $\epsilon$ is large.  This was also present in the original planar model, and is unfortunately exaggerated at small initial radii.   This behavior is analogous to the waiting behavior of nonlinear diffusion fronts described in Kath \cite{waiting}. Curiously, the slowing down of the front  occurs after the derivative of the boundary temperature becomes nonzero. The somewhat counter-intuitive behavior of the heat front slowing down when the drive temperature is increasing is also predicted in the numerical model though to a much smaller degree, so this effect appears to be genuine, albeit exaggerated greatly in our approximate model. 



\subsection{Temperature Profiles}
\begin{figure}
    \centering
    \begin{subfigure}{0.49\textwidth}
        \includegraphics[width=\linewidth]
        {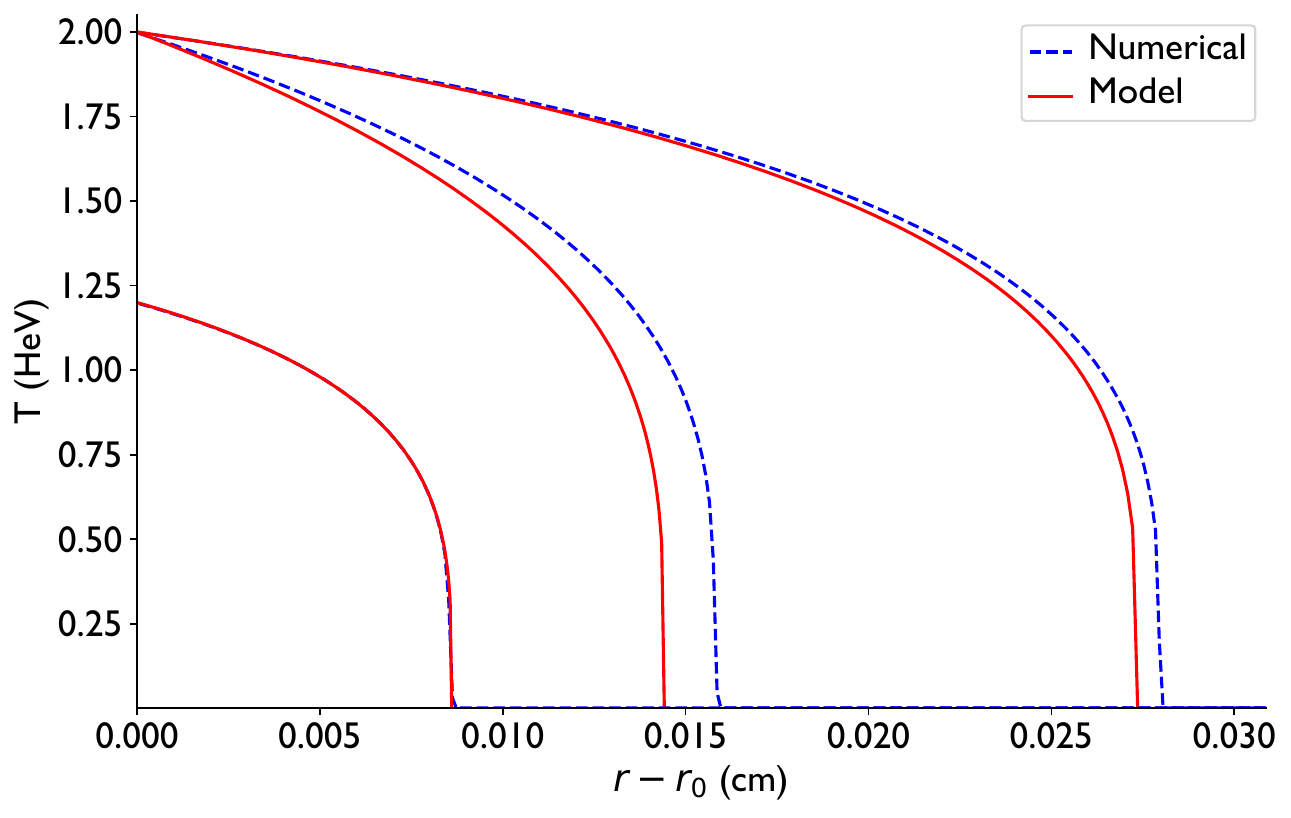}
        \caption{$r_0 = 1000$ cm}
    \end{subfigure}
    \hfill
    \begin{subfigure}{0.49\textwidth}
    \includegraphics[width=\linewidth]{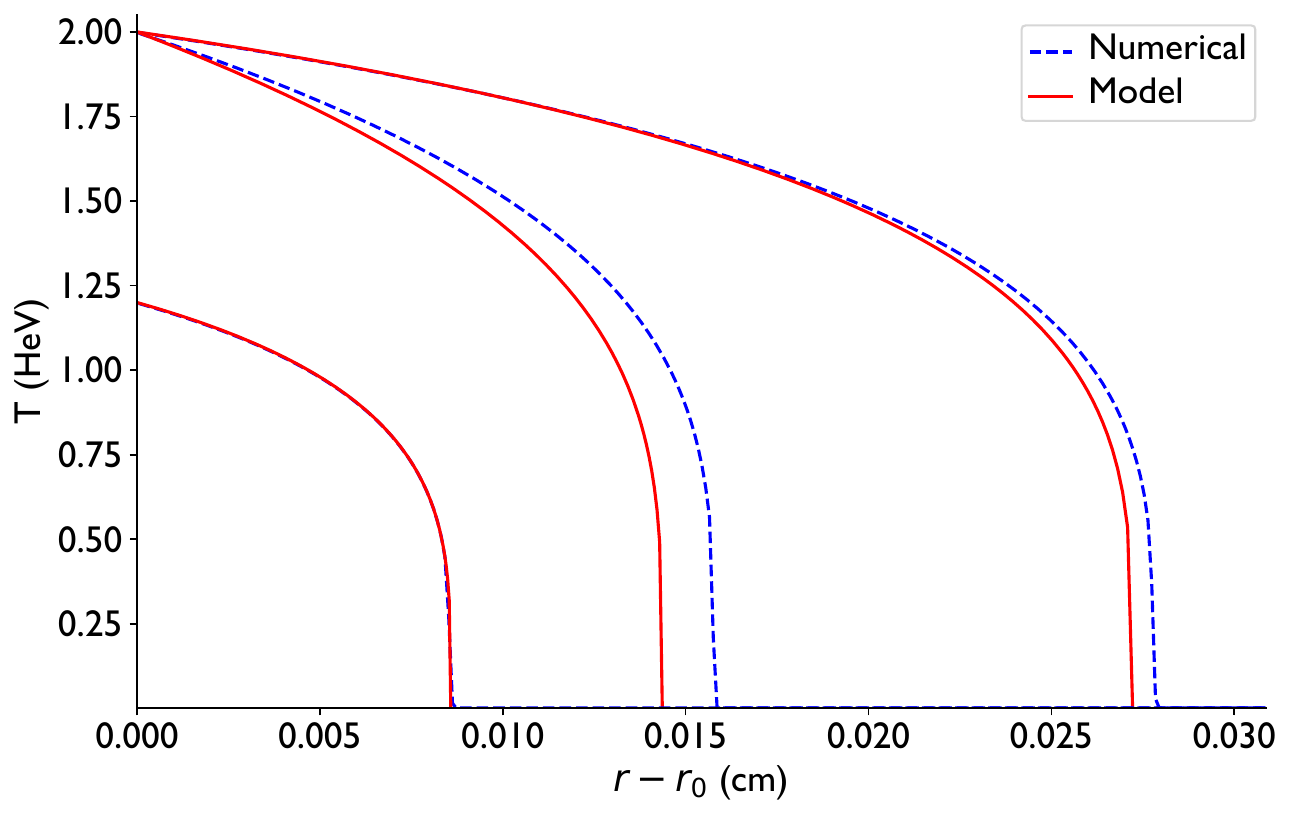}
        \caption{$r_0 = 1$ cm}
        
    \end{subfigure}

    \vspace{5mm} 
    \begin{subfigure}{0.49\textwidth}
        \includegraphics[width=\linewidth]{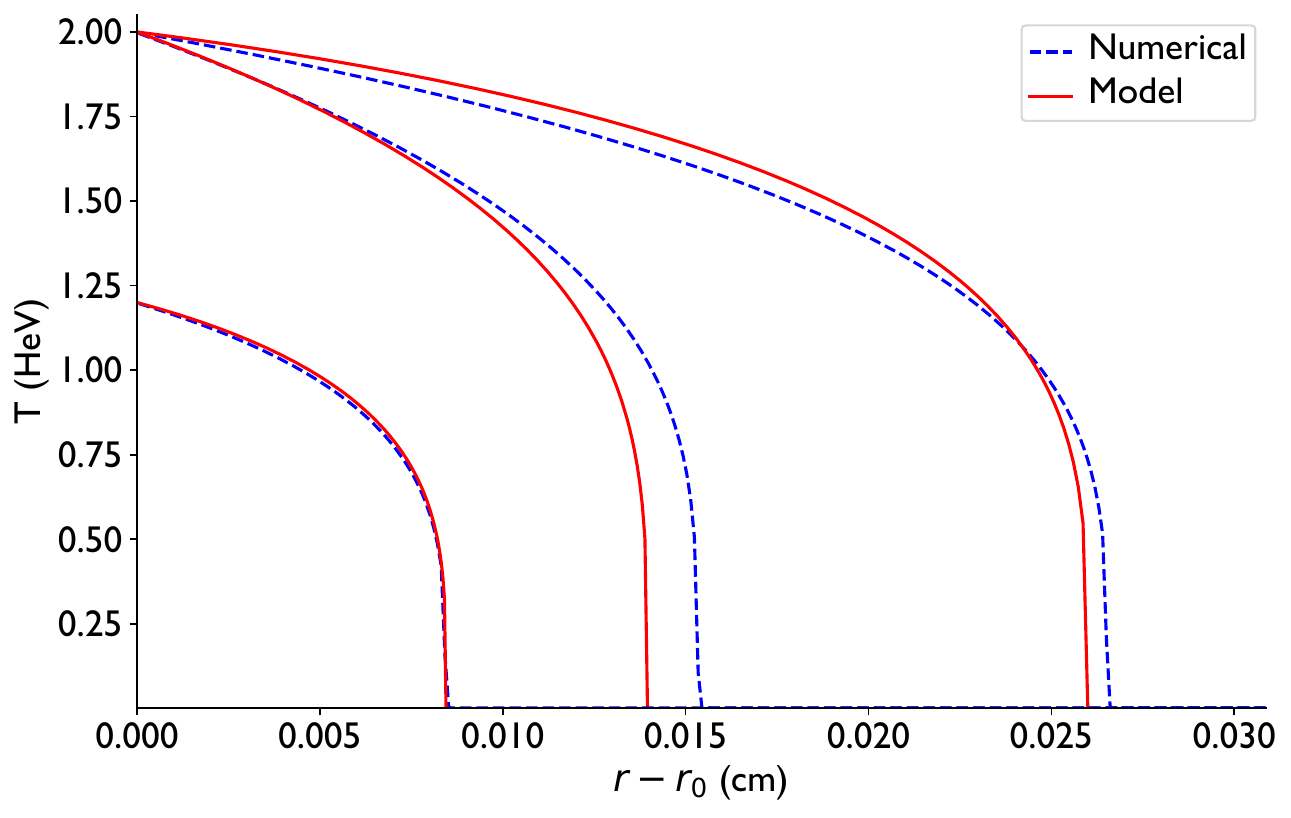}
        \caption{$r_0 = 0.1$ cm}
        
    \end{subfigure}
    \hfill
    \begin{subfigure}{0.49\textwidth}
        \includegraphics[width=\linewidth]{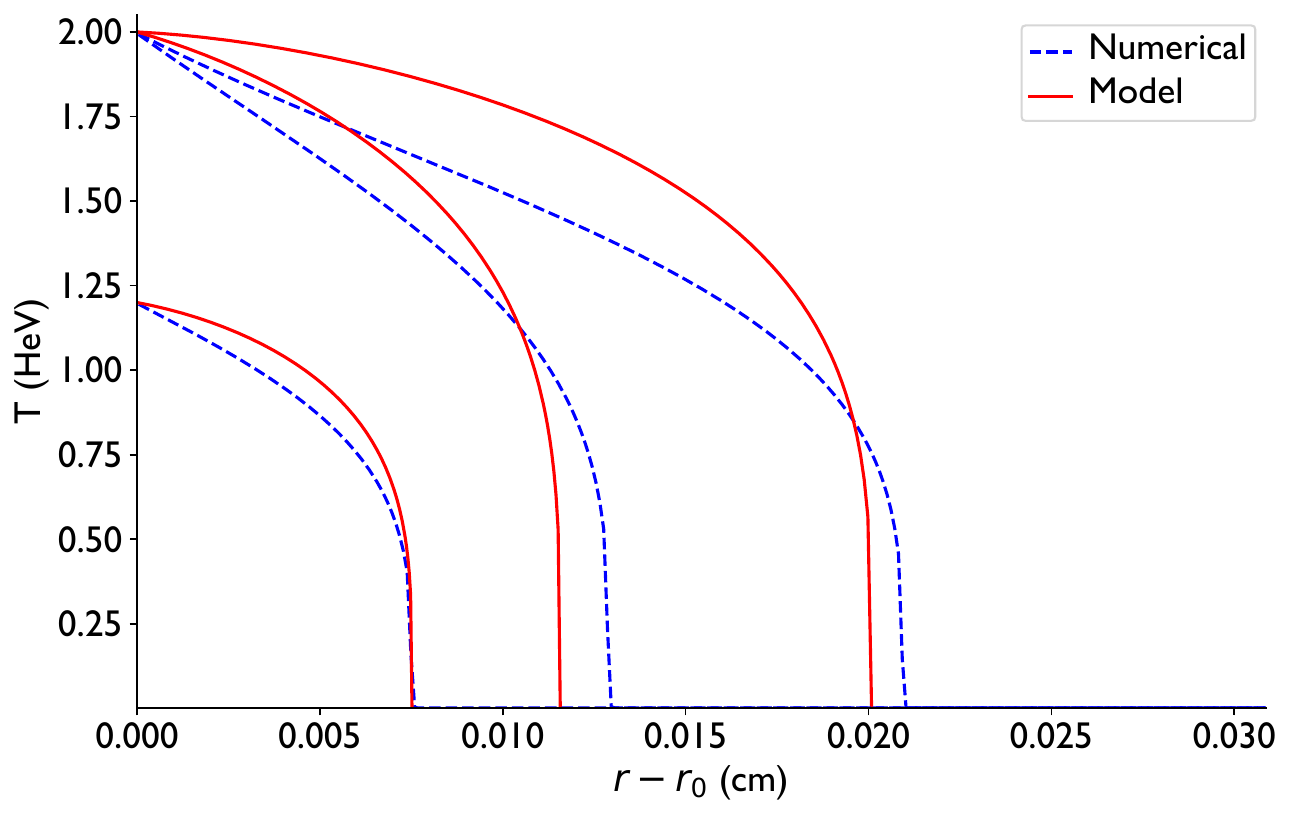}
        \caption{$r_0 = 0.01$ cm}
    \end{subfigure}
    
    \caption{Comparison of the numerical (dashed) and model (solid) temperature profiles in gold at several different inner radii in spherical geometry.}
    \label{fig:gold}
\end{figure}

\begin{figure}
    \centering
    \begin{subfigure}{0.49\textwidth}
        \includegraphics[width=\linewidth]{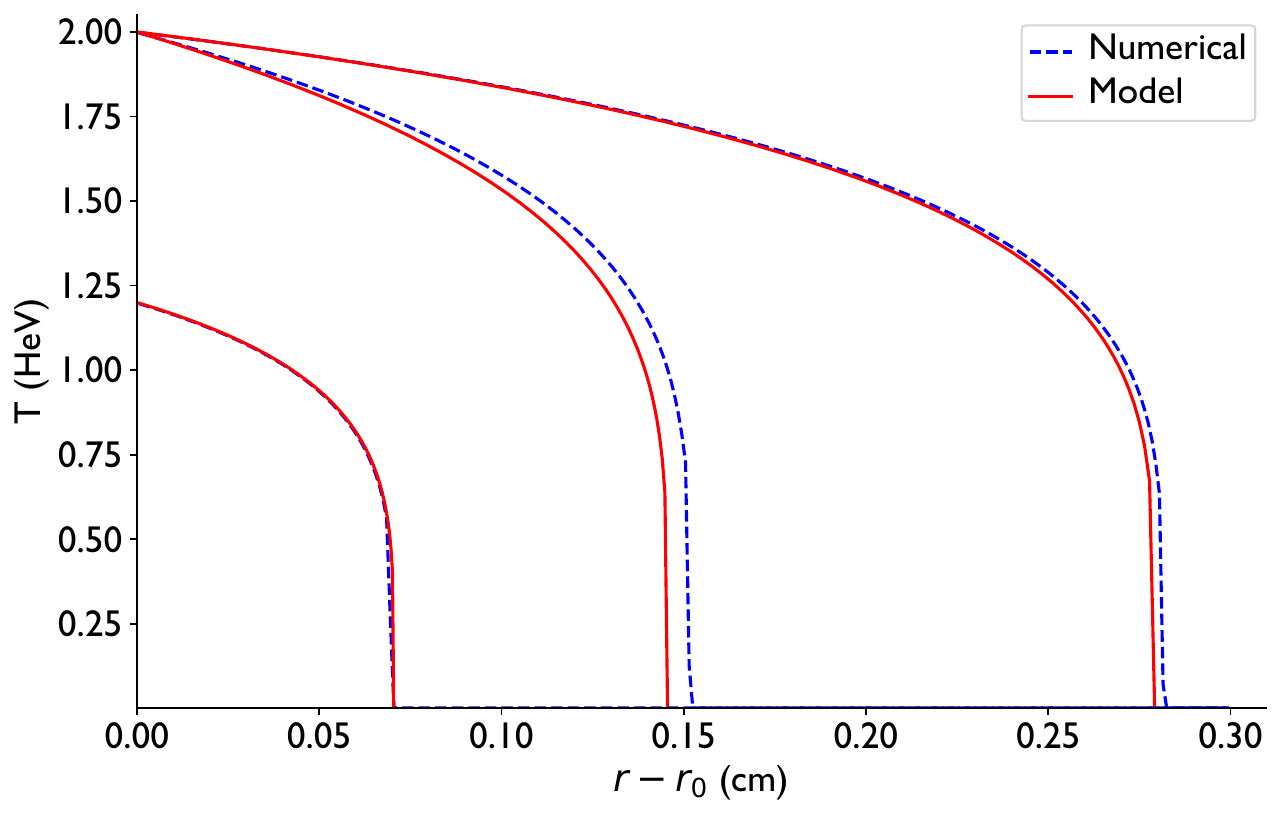}
        \caption{$r_0 = 1000$ cm}
    \end{subfigure}
    \hfill
    \begin{subfigure}{0.49\textwidth}
    \includegraphics[width=\linewidth]{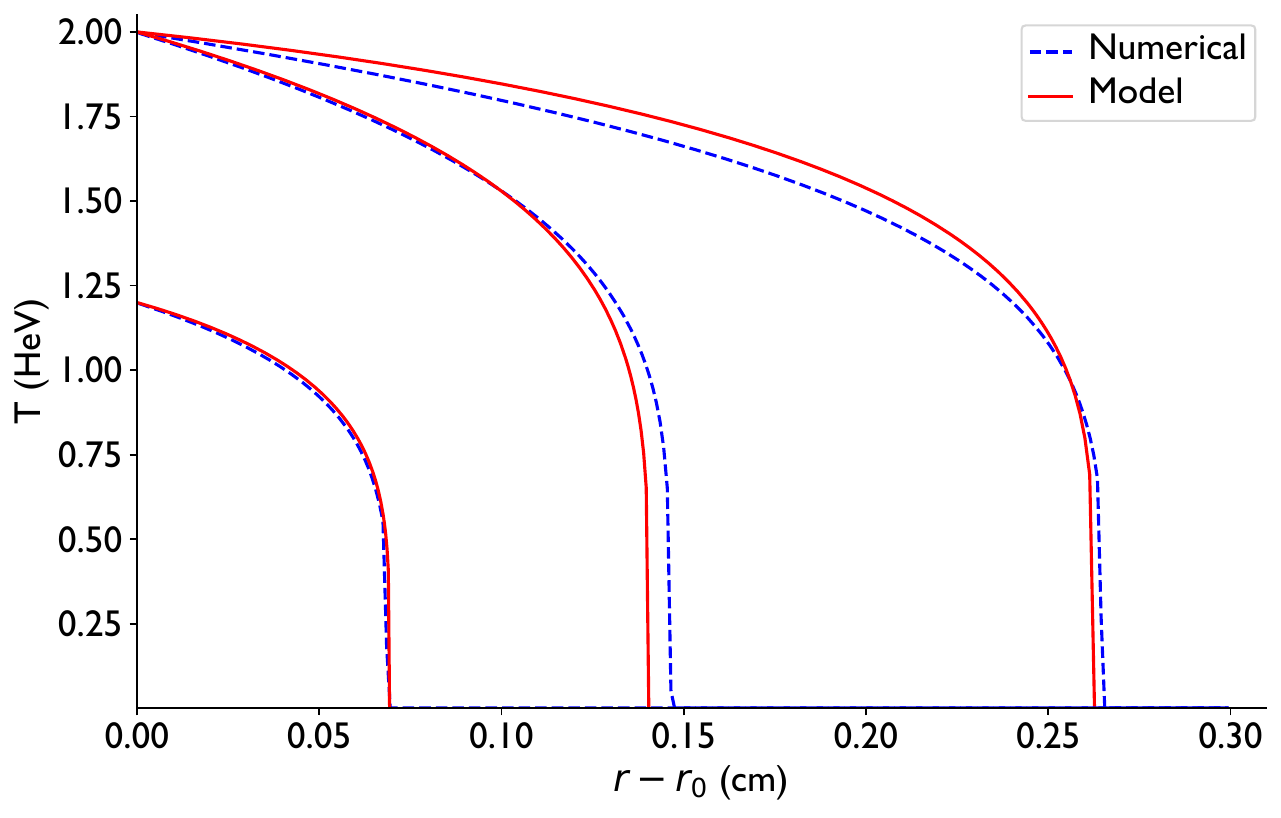}
        \caption{$r_0 = 1$ cm}
        
    \end{subfigure}

    \vspace{5mm} 
    \begin{subfigure}{0.49\textwidth}
        \includegraphics[width=\linewidth]{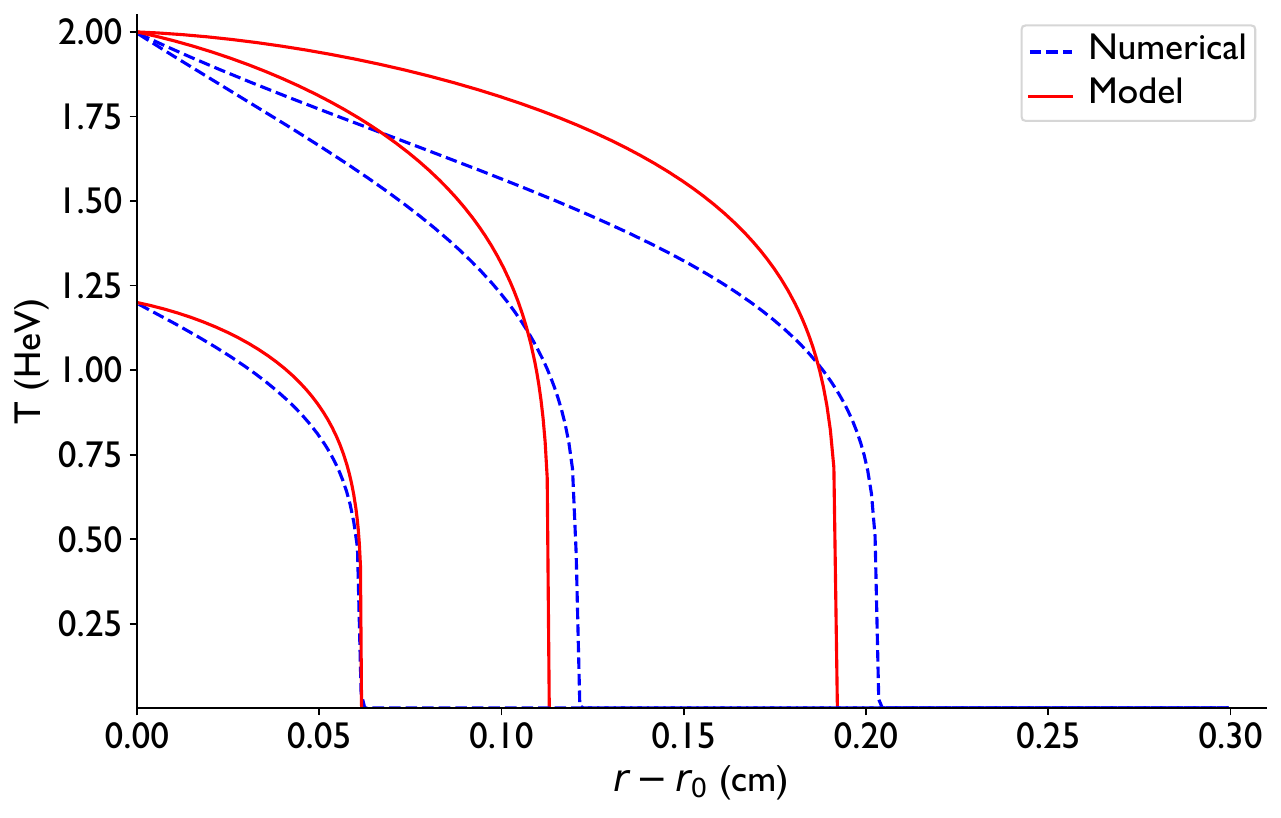}
        \caption{$r_0 = 0.1$ cm}
        
    \end{subfigure}
    \hfill
    \begin{subfigure}{0.49\textwidth}
        \includegraphics[width=\linewidth]{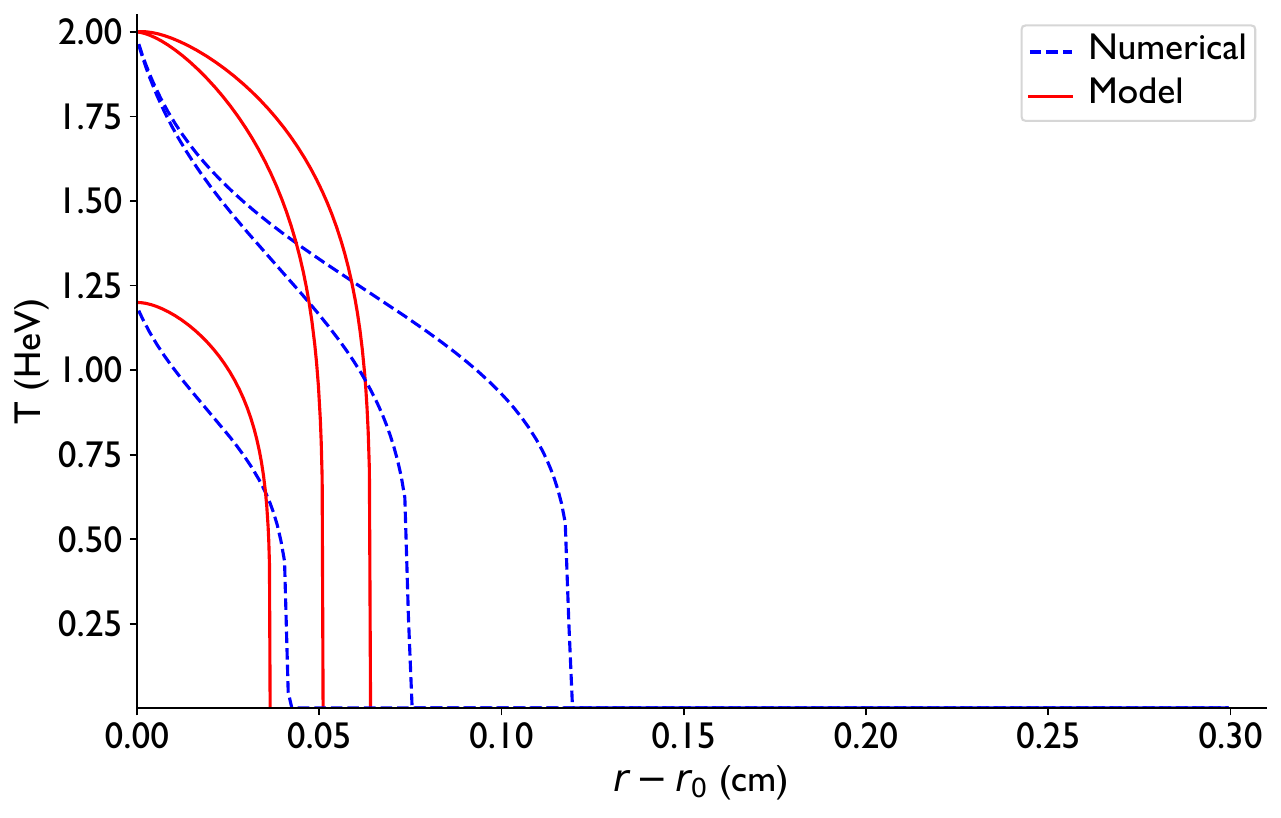}
        \caption{$r_0 = 0.01$ cm}
    \end{subfigure}
    
    \caption{Comparison of the numerical (dashed) and model (solid) temperature profiles in SiO$_2$ at several different inner radii in spherical geometry.}
    \label{fig:SiO2}
\end{figure}

Beyond the wavefront position, we are also interested in the predicted temperature profile in the material. We plot the spherical geometry temperature profile at 1, 2, and 3 ns for gold in Figure \ref{fig:gold} and SiO$_2$ in Figure \ref{fig:SiO2}. In these figures we can see the genesis of the mismatch of the solution from our model at small radii. As the inner radius decreases to $r_0 = 0.01$ cm, panel (d) in both figures, the assumption that the derivative of the temperature is small is violated in the numerical solutions. The sharp change in the solution near $r_0$ results in a wave moving faster than the model predicts. As mentioned above, the large values of $\epsilon$ seem to have smaller geometric effects in that the solution in SiO$_2$ has a much sharper change near $r_0$ than gold.


\begin{figure}
    \centering
    \begin{subfigure}{0.49\textwidth}
        \includegraphics[width=\linewidth]
        {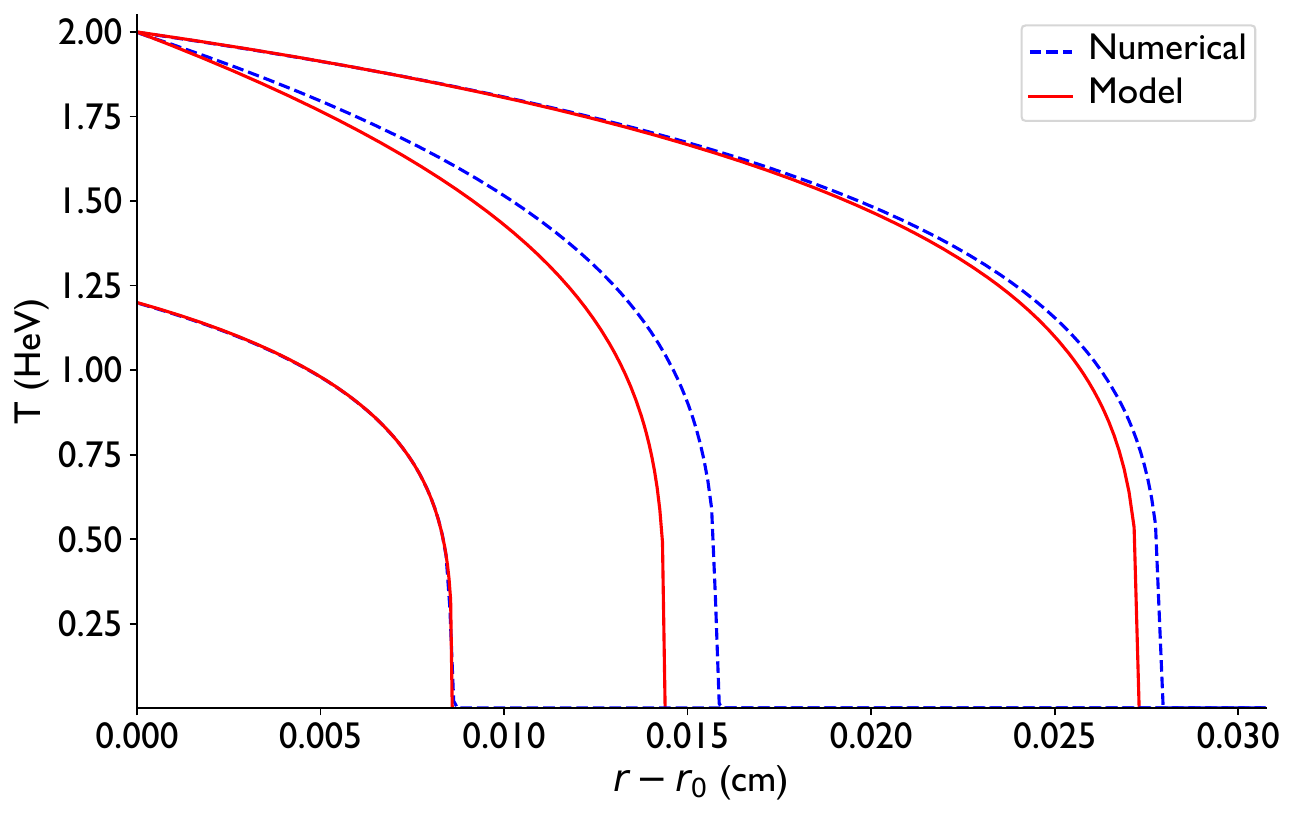}
        \caption{$r_0 = 1$ cm}
    \end{subfigure}
    \hfill
    \begin{subfigure}{0.49\textwidth}
    \includegraphics[width=\linewidth]{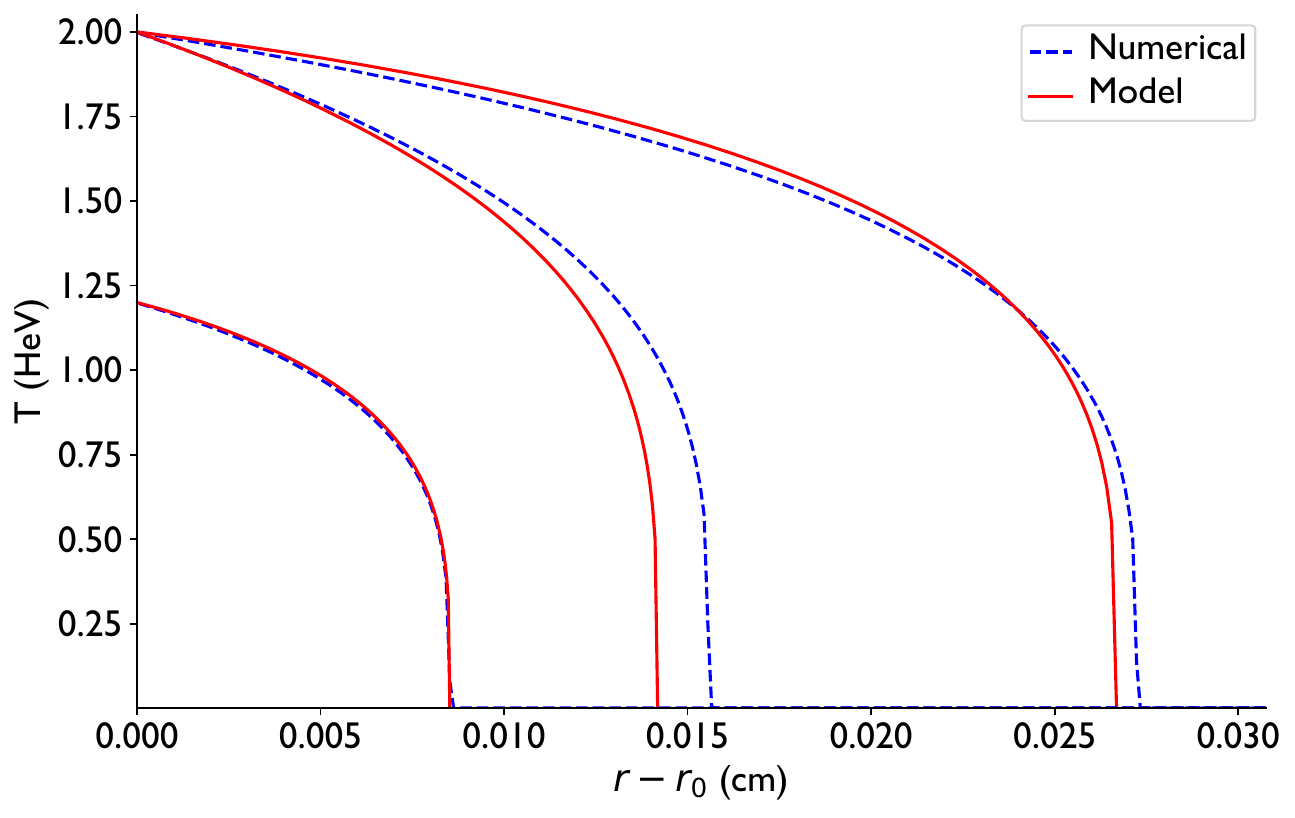}
        \caption{$r_0 = 0.1$ cm}
        
    \end{subfigure}

    \vspace{5mm} 
    \begin{subfigure}{0.49\textwidth}
        \includegraphics[width=\linewidth]{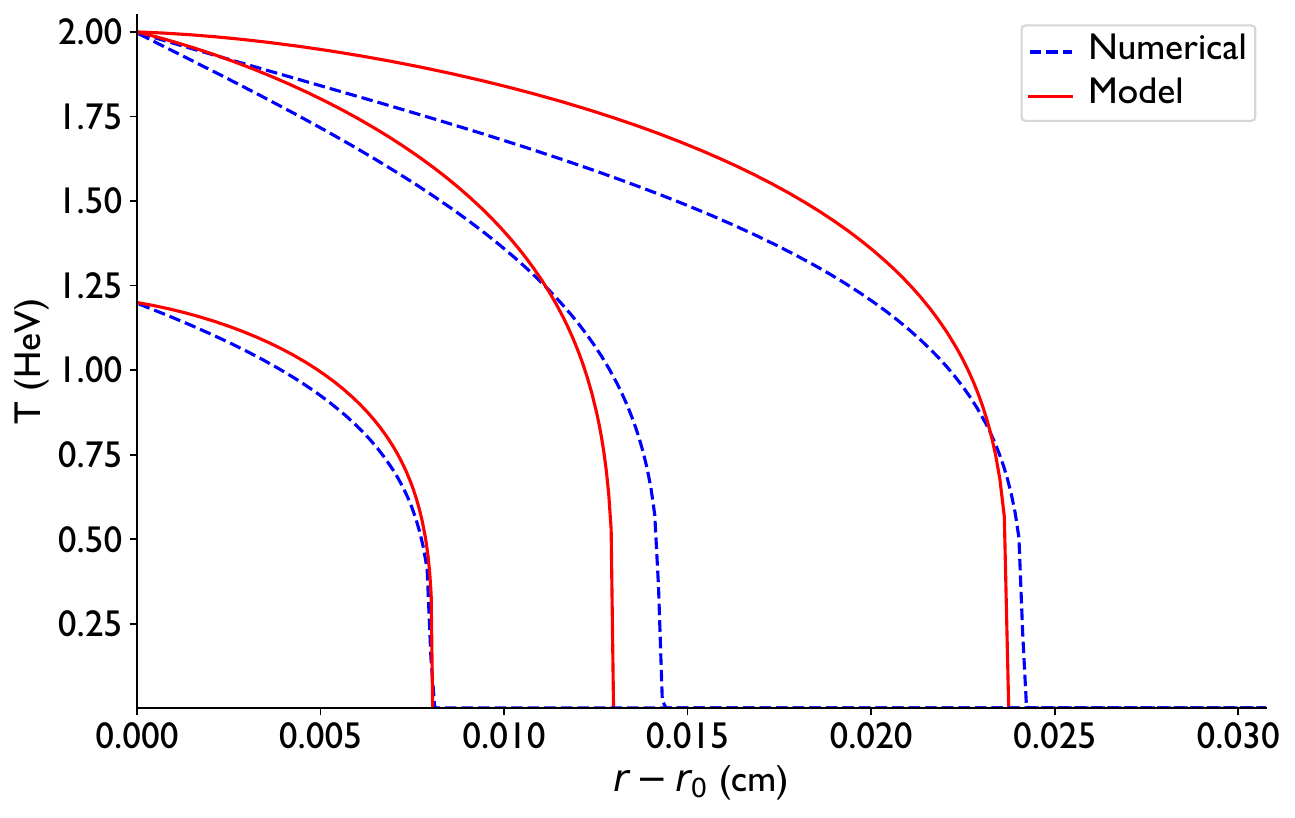}
        \caption{$r_0 = 0.01$ cm}
        
    \end{subfigure}
    \hfill
    \begin{subfigure}{0.49\textwidth}
        \includegraphics[width=\linewidth]{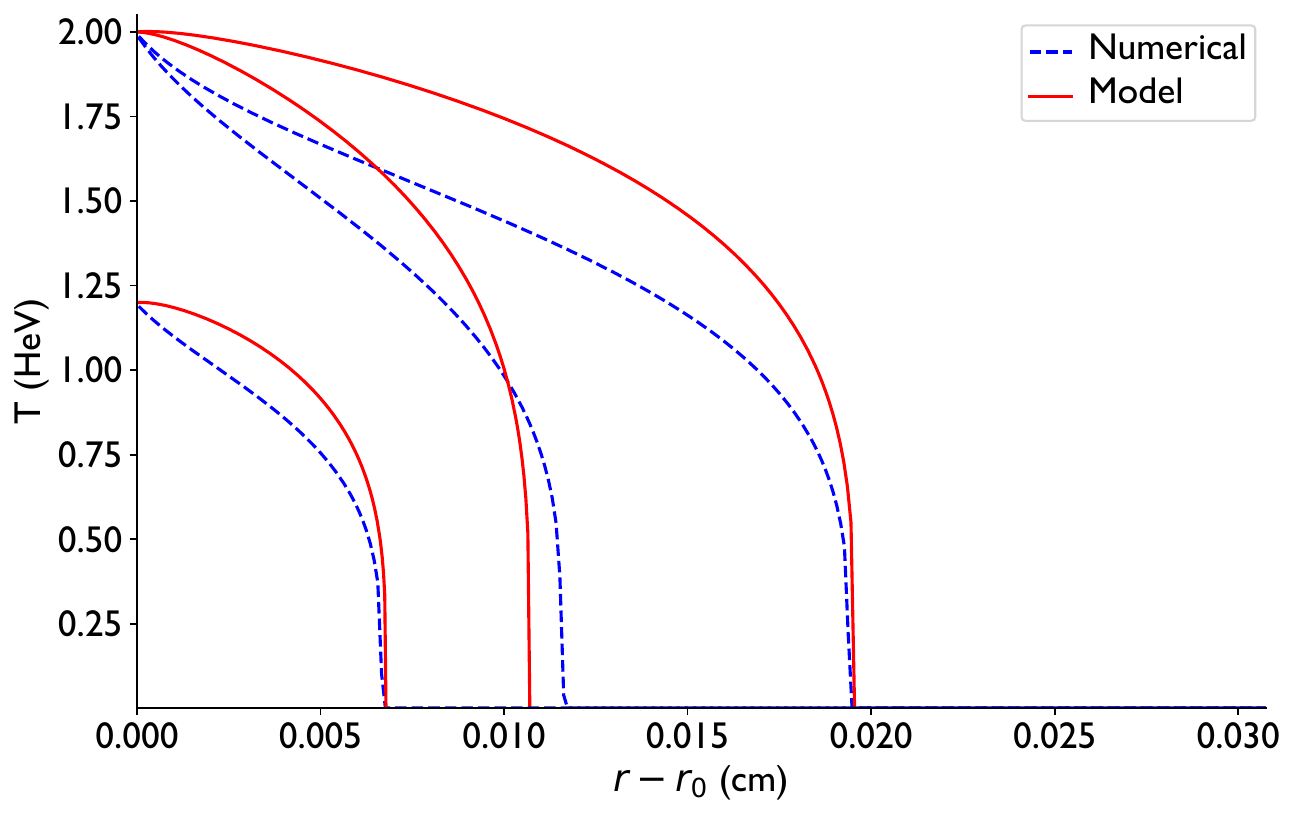}
        \caption{$r_0 = 0.001$ cm}
    \end{subfigure}
    
    \caption{Comparison of the numerical (dashed) and model (solid) temperature profiles in gold at several different inner radii in cylindrical geometry.}
    \label{fig:goldc}
\end{figure}
\begin{figure}
    \centering
    \begin{subfigure}{0.49\textwidth}
        \includegraphics[width=\linewidth]{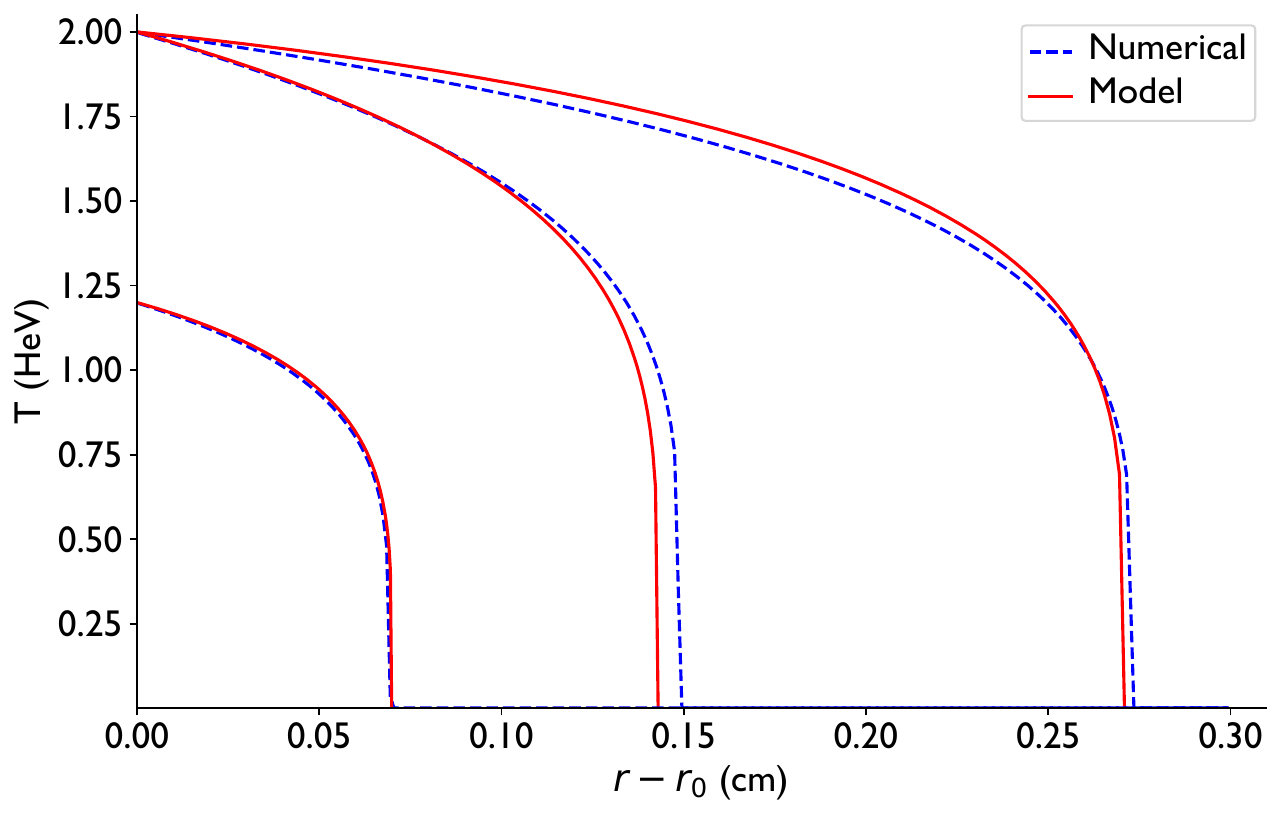}
        \caption{$r_0 = 1$ cm}
    \end{subfigure}
    \hfill
    \begin{subfigure}{0.49\textwidth}
    \includegraphics[width=\linewidth]{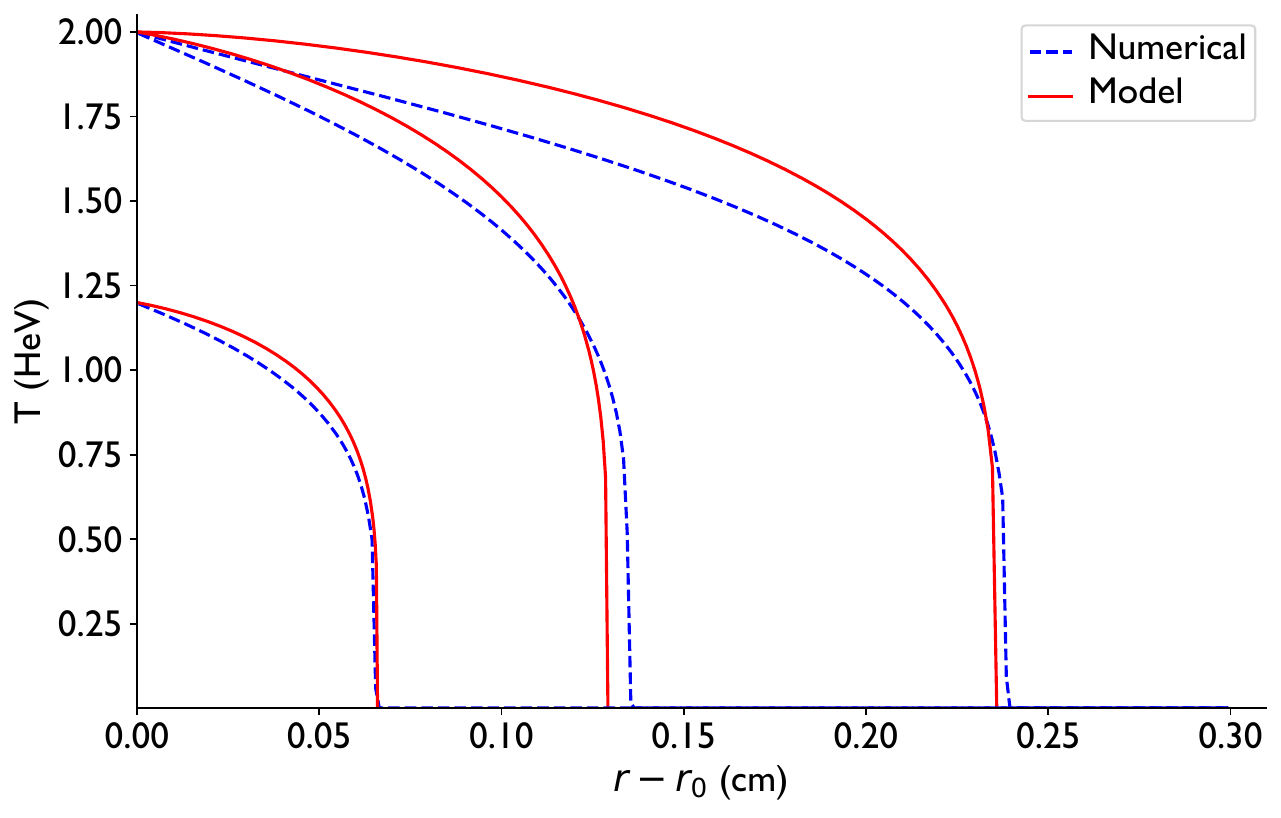}
        \caption{$r_0 = 0.1$ cm}
        
    \end{subfigure}

    \vspace{5mm} 
    \begin{subfigure}{0.49\textwidth}
        \includegraphics[width=\linewidth]{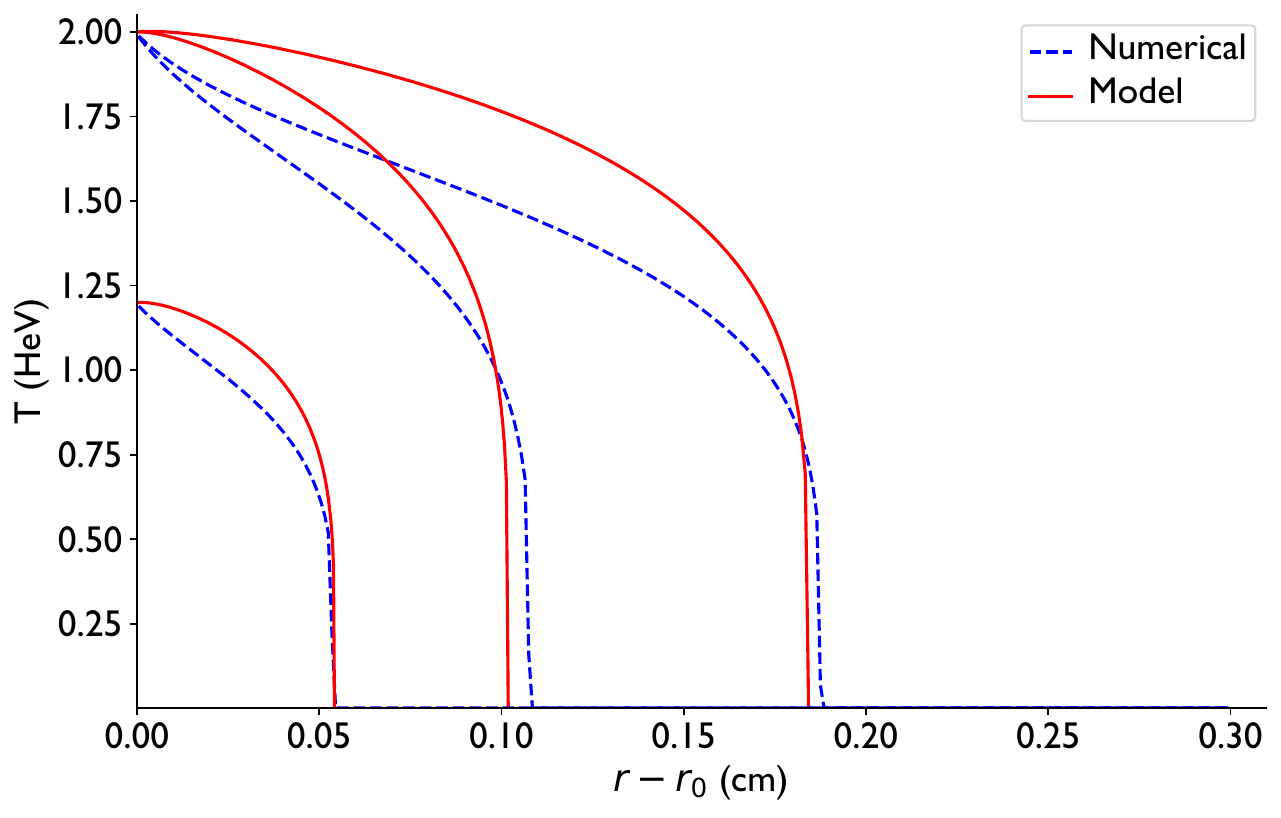}
        \caption{$r_0 = 0.01$ cm}
        
    \end{subfigure}
    \hfill
    \begin{subfigure}{0.49\textwidth}
        \includegraphics[width=\linewidth]{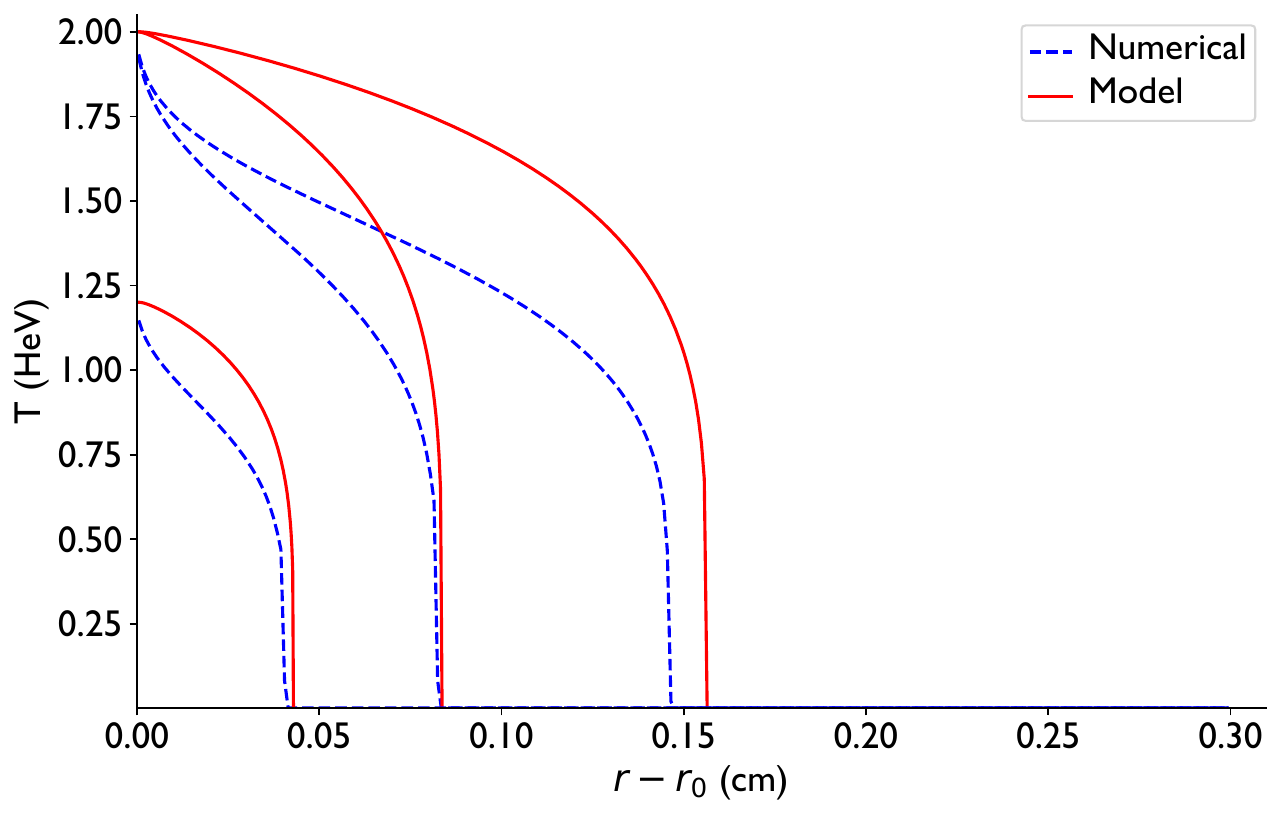}
        \caption{$r_0 = 0.001$ cm}
    \end{subfigure}
    
    \caption{Comparison of the numerical (dashed) and model (solid) temperature profiles in SiO$_2$ at several different inner radii in cylindrical geometry.}
    \label{fig:SiO2c}
\end{figure}
The temperature profiles in cylindrical geometry are shown in Figures \ref{fig:goldc} and \ref{fig:SiO2c}. In this case we see similar effects of geometry as in the spherical case, but it takes smaller values of $r_0$ before the numerical solution has a gradient near the left boundary that is not captured by our model.

\section{Conclusions and future work}

This extension to the Hammer and Rosen model has resulted in a significant increase in the domain of physical systems available to be approached by semi-analytic methods. Of particular interest currently are the small spherical systems developed during inertial confinement fusion implosions. This model will be a boon for rapidly iterating on ICF capsule designs, or for other systems which radiate heat in a curved geometry, due to the arbitrary boundary temperature profile allowed by the method.

Potential future work includes the extension of the model to converging heat waves. A model for such a scenario is complicated by the fact that the problem can no longer be cast as semi-infinite. Additionally, our temperature profiles for small initial radii indicate that in this limit, the assumption that the temperature gradient away from the wavefront is negligible is no longer true. Therefore, a more complex model that allows for a non-neglible temperature gradient near the boundary may be able to capture effects at small radius.

\newpage

\vspace{0.5cm}

\newpage

\bibliography{aipsamp}

\end{document}